\documentclass[letterpaper, 10 pt, conference]{ieeeconf}

\IEEEoverridecommandlockouts 
\makeatletter

\let\proof\@undefined
\let\endproof\@undefined
\makeatother
\let\labelindent\relax

\usepackage{jhoagg}
\usepackage{mathtools}
\usepackage{amsfonts}
\usepackage{amssymb,latexsym}
\usepackage[psamsfonts]{eucal}
\usepackage{amsthm}
\usepackage{graphicx}
\usepackage[inline]{enumitem}
\usepackage{indentfirst}
\usepackage{setspace}
\usepackage{microtype}
\usepackage{threeparttable}
\usepackage{float}
\usepackage{cleveref}
\usepackage{afterpage}
\usepackage{placeins}
\usepackage{cancel}
\usepackage{cite}
\usepackage{leftidx}
\usepackage{breqn}
\usepackage[export]{adjustbox}
\usepackage{comment}
\usepackage[ruled, linesnumbered, lined]{algorithm2e}
\usepackage[dvipsnames]{xcolor}
\usepackage{xcolor}
\usepackage{soul}
\usepackage{siunitx}
\usepackage{subcaption}
\DeclareMathOperator*{\argmax}{argmax} 
\DeclareMathOperator*{\argmin}{argmin} 

\usepackage{tikz}
\usepackage{pgf}
\usepackage{pgfplots}
\usepgflibrary{plothandlers}
\usepgfplotslibrary{external} 
\pgfkeys{/pgf/number format/.cd,1000 sep={}}  
\usetikzlibrary{calc,shapes.misc,plotmarks}
\pgfplotsset{compat=1.13}

\let\originalleft\left
\let\originalright\right
\renewcommand{\left}{\mathopen{}\mathclose\bgroup\originalleft}
\renewcommand{\right}{\aftergroup\egroup\originalright}

\newcounter{thm} 
\newtheorem{theorem}[thm]{\indent Theorem}

\newtheorem{assumption}{\indent Assumption}

\newtheorem{proposition}{\indent Proposition}

\newtheorem{lemma}{\indent Lemma}

\newtheorem{corollary}{\indent Corollary}

\newtheorem{definition}{\indent Definition}

\newtheorem{remark}{\indent Remark}

\newtheorem{example}{\indent Example}

\newtheorem{Simulation}{Simulation}

\newtheorem{fact}{\indent Fact}

\newtheorem{conjecture}{\indent Conjecture}

\newtheorem{experiment}{\indent Experiment}

\allowdisplaybreaks

\renewcommand{\theenumi}{{\it (\alph{enumi})}}

\usepackage{cite}
\usepackage{accents}

\newlength\figureheight 
\newlength\figurewidth

\allowdisplaybreaks

\graphicspath{ {Figures/} }
\DeclareGraphicsExtensions{.png}

\DeclareMathAlphabet{\mathcal}{OMS}{cmsy}{m}{n} 

\crefname{equation}{}{}

\usepackage[ruled, linesnumbered, lined]{algorithm2e}
\SetKwInput{KwInit}{Initialize}
\SetKwFunction{SafeSet}{SafeSet}
\SetKwFunction{GetSafe}{GetSafe}
\SetKwFunction{GetUnsafe}{GetUnsafe}

\SetCommentSty{mycommfont}

\begin{document}
	\title{Control Barrier Functions With Real-Time Gaussian Process Modeling}
	
	\author{Ricardo Gutierrez and Jesse B. Hoagg
		\thanks{R. Gutierrez and J. B. Hoagg are with the Department of Mechanical and Aerospace Engineering, University of Kentucky, Lexington, KY, USA. (e-mail: Ricardo.Gutierrez@uky.edu, jesse.hoagg@uky.edu).}
		\thanks{R. Gutierrez is supported by the Fulbright-SENACYT Scholarship. 
        This work is also supported in part by USDA NIFA (2024-69014-42393) and AFOSR (FA9550-20-1-0028).
        }
	}
	\maketitle
	
\begin{abstract}
We present an approach for satisfying state constraints in systems with nonparametric uncertainty by estimating this uncertainty with a real-time-update Gaussian process (GP) model.
Notably, new data is incorporated into the model in real time as it is obtained and select old data is removed from the model. 
This update process helps improve the model estimate while keeping the model size (memory required) and computational complexity fixed.
We present a recursive formulation for the model update, which reduces time complexity of the update from $\mathcal{O}(p^{3})$ to $\mathcal{O}(p^{2})$, where $p$ is the number of data used.
The GP model includes a computable upper bound on the model error.
Together, the model and upper bound are used to construct a control-barrier-function (CBF) constraint that guarantees state constraints are satisfied. 
\end{abstract}
	
	
	
\section{Introduction}
%
Control systems are often required to achieve performance goals (e.g., formation control, locomotion, destination seeking) while satisfying state constraints. 
These objectives can sometimes result in competing/conflicting requirements, which can be addressed with approaches such as model predictive control (e.g., \cite{borrelli2017predictive,bemporad2002model,tondel2003algorithm}) and barrier functions (e.g.,~\cite{prajna2007framework,wieland2007constructive,ames2016control,nguyen2016exponential,xiao2021high,tan2021high}).

Control barrier functions (CBFs) ensure satisfaction of state constraints by achieving forward invariance of a safe set \cite{wieland2007constructive}. 
CBFs are often implemented as constraints of quadratic programs, while a performance-based cost is minimized \cite{ames2016control}. 
However, CBF are model based and thus, susceptible to model uncertainties, which can lead to degraded performance or state-constraint violations \cite{gutierrez2024adaptive}.
Robust CBFs can be effective for constraint satisfaction with model uncertainties \cite{jankovic2018robust}. 
However, this approach relies on worse-case bounds, which may be too conservative. 
Furthermore, robust CBFs do not directly address the effect of model uncertainty on the desired control, which can lead to performance degradation.

Adaptive CBFs have been implemented to address safety in systems with parametric uncertainties (e.g., \cite{cohen2022high,taylor2020adaptive, gutierrez2024adaptive}) using methods such as recursive-least-squares estimation and providing guarantees of stability and estimation-error bounds. 
CBF has also been studied together with parametric uncertainties using set-membership identification \cite{lopez2020robust, cohen2022robust}.
Although adaptive parametric CBFs can be effective, their applicability is limited to parametric uncertainty \cite{boffi2022nonparametric}.

Gaussian processes (GP) regression is commonly used to model nonparametric uncertainty from data. 
Under the assumption that the unknown function belongs to a reproducing kernel Hilbert space (RKHS), the GP estimation error can be upper bounded \cite{hashimoto2022learning,reed2024error,jagtap2020control,cheng2020safe}.
GPs have been implemented together with CBFs to ensure state constraint satisfaction with nonparametric uncertainties \cite{jagtap2020control,cheng2020safe,khan2020gaussian,khan2021safety,reed2024error}.
However, \cite{jagtap2020control,cheng2020safe,khan2020gaussian,khan2021safety,reed2024error} focus on static datasets. 
In many control applications, data are collected in real time. 
To address this, \cite{lederer2021gaussian} proposed a streaming GP regression approach based on grids and locally growing random trees. 
However, this method remains susceptible to data overflow and growing model complexity. 
Thus, there is need for a real-time GP-model approach that can efficiently handle continuous data streams without an explosion in model size and/or computational complexity.

This article presents an approach for updating a Gaussian process (GP) model in real time using streaming data. 
Notably, new data is incorporated into the model as it is obtained and carefully selected old data is removed from the model. 
The model update process improves the local model estimate and maintains the best-possible nonlocal model, while keeping the number of data used in the model fixed, which keeps the required memory and computational complexity fixed.
We also present a recursive formulation for the model update, which reduces time complexity of the update from $\mathcal{O}(p^{3})$ in the batch algorithm to $\mathcal{O}(p^{2})$ in the recursive algorithm, where $p$ is the number of data used.
We note that the GP model includes a computable upper bound on the model error \cite{hashimoto2022learning}, which is used along with the GP model to construct a CBF constraint. 
Then, we present a minimum-intervention quadratic-program-based control that guarantees state constraint satisfaction. 
Note that the GP model estimate can also be effectively incorporated into a desired control law.
This feature can help to improve performance while ensuring state-constraint satisfaction.
We demonstrate state constraint satisfaction and performance using 2 examples: a nonlinear pendulum and a nonholonomic robot.



\section{Notation}
Let $q:\mathbb{R}^{n} \rightarrow \mathbb{R}$ be continuously differentiable. 
Then, $q^\prime:\mathbb{R}^{n} \rightarrow \mathbb{R}^{1\times n}$ is defined by $q^\prime(x) \triangleq \textstyle \frac{\partial q(x)}{\partial x}$. 
The Lie derivative of $q$ along the vector fields of $\upsilon:\mathbb{R}^{n} \rightarrow \mathbb{R}^{n \times l}$ is defined as $L_\upsilon q(x) \triangleq q^\prime(x) \upsilon(x)$.
If $l=1$, then for all positive integers $r$,  define $L_\upsilon^{r}q(x) \triangleq L_{\upsilon} L_{\upsilon}^{r-1} q(x)$.

Let $B\in \mathbb{R}^{m \times n}$. Then, $(B)_{[i]}$ denotes the $i$th row of $B$, $(B)_{[\lnot i]}$ denotes the matrix $B$ obtained by removing the $i$th row, $(B)_{[i,j]}$ denotes the $(i,j)$-th entry of $B$, $(B)_{[i, \lnot j]}$ denotes the $i$th row of $B$ removing the $j$th column and $(B)_{[\lnot i, j]}$ denotes the $j$th column of B removing the $i$th row.
Let $C\in \mathbb{R}^{l}$. Then, $(C)_{[i]}$ denotes the $i$th entry of $C$ and $(C)_{[\lnot i]}$ denotes the vector $C$ obtained by removing the $i$th entry.

The operator $\big|\cdot\big|$ denotes the absolute value of its argument applied element-wise. Let $1_{l} \in \mathbb{R}^{l}$ denotes a vector of ones.
 

\section{Problem Formulation}
Consider the dynamic system
\begin{equation}
		\Dot{x}(t)=f(x(t))+w(x(t))+g(x(t))u(t),
		\label{eq:dyn}
\end{equation}
where $x(t) \in \mathbb{R}^n$ is the state; $x(0)=x_{0} \in \BBR^n$ is the initial condition; $u: [0, \infty) \rightarrow \mathbb{R}^m$ is the control; $f: \mathbb{R}^n \rightarrow \mathbb{R}^{n}$, $g: \mathbb{R}^n \rightarrow \mathbb{R}^{n \times m}$, and $w: \mathbb{R}^{n} \rightarrow \mathbb{R}^{n}$ are locally Lipshchitz continuous on $\mathbb{R}^{n}$. The functions $f$ and $g$ are known, and $w$ is unknown.

Let $\psi_{0}:\mathbb{R}^{n} \rightarrow \mathbb{R}$ be continuously differentiable, and define the \textit{safe set}
    \begin{equation}
	C_{0} \triangleq \{x \mid \psi_{0}(x)\geq0\},
	\label{set:safe_set}
    \end{equation}
which is the set of states that satisfy the state constraint. 
We make the following assumption:

\begin{assumption}{\rm 
There exists a positive integer $d$ such that for all $i\in\{0,1,\ldots,d-2\}$ and all $x \in \BBR^n$, $L_{g}L_{f}^{i}\psi_{0}(x)=0$ and               $L_{w}L_{f}^{i}\psi_{0}(x)=0$.
}\label{assum:input}
\end{assumption}

Assumption~\ref{assum:input} implies that the relative degree of $\psi_0$ with respect to $u$ and the unknown dynamics $w$ is at least $d$ on $\BBR^n$. 
Thus, we use a higher-order approach to construct a candidate CBF. 
For all $i \in \{1,\ldots,d-1\}$, let $\alpha_{i-1}: \mathbb{R} \rightarrow \mathbb{R}$ be a $(d-i)$-times continuously differentiable extended class-$\SK$ function, and define $\psi_{i}(x) \triangleq L_{f}\psi_{i-1}(x)+\alpha_{i-1} \big( \psi_{i-1}(x) \big)$.
For all $i \in \{1,\ldots,d-1\}$, the zero-superlevel set of $\psi_i$ is given by $ C_{i} \triangleq \{x \mid \psi_{i}(x)\geq0\}$.
We make the following assumption:

\begin{assumption}\label{assum:assump2}
\rm{For all $x \in \mathrm{bd} \hspace{1mm} C_{d-1}$, $L_{g}L_{f}^{d-1}\psi_{0}(x) \neq 0$.} 
\end{assumption}
Assumption~\ref{assum:assump2} implies that $\psi_0$ has relative degree $d$ with respect to $u$ on $\mbox{bd } C_{d-1}$. 
Thus, $\bar C \triangleq \bigcap_{i=0}^{d-1} C_{i}$,
is control forward invariant. 

Next, consider the desired control $u_{\rm{d}}:\mathbb{R}^{n} \times \mathbb{R}^{n} \rightarrow \mathbb{R}^{m}$, which is designed to satisfy performance requirements but may not satisfy the state constraint that  $x(t) \in C_{0}$ for all $t$. The desired control $u_{\rm{d}}(x,\hat{w})$ is a function of the state $x$ and can be parameterized using $\hat{w} \in \mathbb{R}^{n}$, which represents an estimate of the unknown function $w(x)$. In other words, $u_{\rm{d}}(x,w(x))$ is the ideal desired control.

Consider the cost $J:\mathbb{R}^{n} \times \mathbb{R}^{n} \times \mathbb{R}^{m} \rightarrow \mathbb{R}$ defined by
    \begin{equation}
        J(x,\hat{w},\hat{u})\triangleq \frac{1}{2} \Big ( \hat{u}-u_{\rm{d}}(x,\hat{w}) \Big )^{\rmT} H (x,\hat{w}) \Big (\hat{u}-u_{\rm{d}}(x,\hat{w}) \Big ), 
	\label{eq:main_cost_func}
    \end{equation}
where  $H : \mathbb{R}^{n} \times \mathbb{R}^{n} \rightarrow \BBR^{m \times m}$ is locally Lipschitz, and for all $(x,\hat{w}) \in \mathbb{R}^{n} \times \mathbb{R}^{n}$, $H(x,\hat{w})$ is positive definite.

The objective is twofold. 
First, design a feedback control that for all $x \in \mathbb{R}^{n}$, minimizes \eqref{eq:main_cost_func} subject to the state constraint that $x(t) \in \bar C$ for all $t \ge 0$. 
Second, we aim to estimate $w(x)$ so that the desired control $u_{\rm{d}}(x,\hat{w})$ is close to the ideal desired control $u_{\rm{d}}(x,w(x))$.

\section{GP model and deterministic error bounds} \label{sec:GP-General}
    This section provides a brief review of GP modeling. 
    See \cite{williams2006gaussian} for more information.
    Let $\nu \sim \mathcal{N}(0,\rho^{2})$ be additive noise such that $||\nu||_{\infty} \leq \rho$ where $ \rho \geq 0$.
   For all $j \in \{1,2,...,p\}$, let $x_{j} \in \mathbb{R}^{n}$ and define $y_{j} \triangleq w(x_{j})+\nu$.  Let $X \triangleq [x_{1} \hspace{0.5mm}  ... \hspace{0.5mm}  x_{p}]^{\rm{T}}$ and  $Y \triangleq [y_{1} \hspace{0.5mm} ... \hspace{0.5mm}  y_{p}]^{\rm{T}}$.
    
    Let $q : \mathbb{R}^{n} \times \mathbb{R}^{n} \rightarrow \mathbb{R}$ be symmetric, positive definite and continuosly differentiable. Then, define
    \begin{equation}
        Q(x,X)\triangleq[q(x,x_{1}) \hspace{1mm} ...  \hspace{1mm} q(x,x_{p})]^{\rm{T}},
        \label{eq:GP_Qi}
    \end{equation}
    \begin{equation}
        \Omega(X)\triangleq P(X)+\rho^{2}I_{p} ,
        \label{eq:GP_Omega_def}
    \end{equation}
    \begin{equation}
        P(X)\triangleq
        \left [ \begin{smallmatrix}
            q(x_{1},x_{1}) & \cdots   & q(x_{p},x_{1})  \\
            \vdots           & \ddots   & \vdots            \\
            q(x_{1},x_{p}) & \cdots   & q(x_{p},x_{p})  \\
        \end{smallmatrix} \right ].
        \label{eq:GP_P_def}
    \end{equation}
    
   The predictive mean $\mu : \mathbb{R}^{n} \times \mathbb{R}^{p\times n} \times\mathbb{R}^{p \times n} \rightarrow \mathbb{R}^{n}$ and standard deviation $\sigma  : \mathbb{R}^{n} \times \mathbb{R}^{p\times n} \times \mathbb{R}^{p\times n} \rightarrow \mathbb{R}$ of the unknown function $w(x)$ are given by
    \begin{equation}
        \mu(x,X,Y)\triangleq Y^{\rm{T}}\Omega(X)^{-1}Q(x,X),
        \label{eq:GP_stacked_mean}
    \end{equation}
    \begin{equation}
        \sigma(x,X) \triangleq \sqrt{q(x,x)-Q(x,X)^{\rm{T}}\Omega(X)^{-1}Q(x,X)}.
        \label{eq:GP_stacked_std}
    \end{equation}

    Let $\mathcal{H}_{q}$ be the RKHS associated with the kernel $q$ and let $||\cdot||_{q}$ denotes the RKHS induced norm. The following assumption indicates the existence of a known bound in the RKHS norm of the unknown function $w$. This assumption is standard \cite{hashimoto2022learning,jagtap2020control,reed2024error}.
    \begin{assumption}\rm{
         For all $i\in \{1,2,...,n\}$, $(w)_{[i]} \in \mathcal{H}_{q}$, and there exists $b>0$ such that $||(w)_{[i]}||_{q} \leq b$. 
        }
        \label{assum:RKHS_norm}
    \end{assumption}

    Assumption \ref{assum:RKHS_norm} implies that $(w)_{[i]}$ is the weighted sum of kernels (i.e., $(w)_{[i]}=\sum_{j=1}^{\infty} \gamma_{j} q(\cdot,x_{j})$) with $||w||_{{q}}=\gamma^{\rm{T}}P(X)\gamma$, where $\gamma \triangleq [\gamma_{1} \hspace{1mm} \cdots \hspace{1mm} \gamma_{p}]^{\rm{T}}$ \cite{maddalena2021deterministic}.

    Next, consider $B : \mathbb{R}^{p \times n} \times \mathbb{R}^{p \times n} \rightarrow \mathbb{R}^{n}$ defined by
    \begin{equation}
        B(X,Y) \triangleq \left [ 
        \begin{smallmatrix}
            \sqrt{b^{2}-\big(Y^{\rm{T}}\Omega(X)^{-1}Y\big)_{[1,1]}+p}\\
            \vdots \\
            \sqrt{b^{2}-\big(Y^{\rm{T}}\Omega(X)^{-1}Y\big)_{[n,n]}+p}
        \end{smallmatrix} \right ].
        \label{eq:GP_stacked_B}
    \end{equation}
    The following proposition, given by \cite{hashimoto2022learning}, shows that $B(X,Y)\sigma(x,X)$ provides an upper bound on the estimation error $|\mu(x,X,Y)-w(x)|$.
    \begin{proposition} \rm{
        Assume assumption \ref{assum:RKHS_norm} is satisfied. Then, for all $x \in \mathbb{R}^{n}$ and all $i \in \{1,2,...,n\}$,
        \begin{equation}
            \big|\big(\mu(x,X,Y)-w(x)\big)_{[i]}\big| \leq \big(B(X,Y)\sigma(x,X)\big)_{[i]}.
            \label{eq:unknown_function_limits}
        \end{equation}
        }
        \label{prop:deterministic_error_bound}
    \end{proposition}

    The mean model $\mu(x,X,Y)$, standard deviation $\sigma(x,X)$, and bound factor $B(X,Y)$ given by \cref{eq:GP_stacked_mean,eq:GP_stacked_std,eq:GP_stacked_B} depend on the data $(X,Y)$. The next section introduces a method for updating the data $(X,Y)$ in real time using new measurements.

\section{Real-time GP modeling} 
    This section presents a method for updating a GP model using data obtained in real time (e.g., during real time control and execution). At each model update, the GP model uses exactly $p$ data points. In other words, each time the model is updated, new data is incorporated and certain old data is removed (i.e., forgotten). This add/remove process prevents an explosion of the memory and computational complexity required to compute $\mu$, $\sigma$, and $B$.

    The first subsection presents an approach to updating the data based on two competing objectives: $(i)$ improving the accuracy of the model in a local neighborhood of the data most recently collected (e.g., improving model for instantaneous control); and $(ii)$ preserving the accuracy of the model outside the local neighborhood (e.g., preserving model for future control actions as the system moves in the state space). These competing objective are achieved by partitioning that data into subsets aimed at achieving $(i)$ and $(ii)$.

    The second subsection presents computationally efficient recursive update equations for $\mu$, $\sigma$ and $B$ as data is added and removed. We note that direct computation of $\mu$, $\sigma$ and $B$ from data $(X,Y)$ has computational complexity $\mathcal{O}(p^3)$ because of the computation of $\Omega(X)^{-1}$. Using the recursive update equations in the second subsection, we update $\mu$, $\sigma$, and $B$ with computational complexity $\mathcal{O}(p^{2})$.  
    
    Unless otherwise stated, in the remainder of this section, all expressions with subscript $k$ are for all $k \in \mathbb{N} \triangleq \{0,1,2,...\}$.

    \subsection{Updating Data} \label{subsec:general_alg}
    Let $p_{\rm{l}}$ be a positive integer less that $p$, which indicates the number of data used for $(i)$ improving the local model, and $p_{\rm{g}}\triangleq p-p_{\rm{l}}$ is the number of data used for $(ii)$ preserving the nonlocal (e.g., global) model. Let $T_{\rm{s}} >0$ be the sampling time at which new data is obtained. 
    
    For all $k \in \mathbb{N}$, at time $kT_{\rm{s}}$, we obtain $x_{k} \triangleq x(kT_{\rm{s}})$ and before time $(k+1)T_{\rm{s}}$, we obtain $y_{k} \triangleq y(kT_{\rm{s}})$. 
    
    This subsection presents a processes for updating the data set, specifically, a method for obtaining $X_{k+1}\in \mathbb{R}^{p \times n}$ and $Y_{k+1} \in \mathbb{R}^{p \times n}$ from $X_{k}$, $Y_{k}$, $x_{k}$ and $y_{k}$.
    Let $X_{0}\in \mathbb{R}^{p \times n}$ and $Y_{0}\in \mathbb{R}^{p \times n}$ be the initial data. Furthermore, let $c_{0}\in \mathbb{R}^{p}$ be a vector, where $p_{\rm{l}}$ entries are zero and $p_{\rm{g}}$ are one.

    For each $k \in \mathbb{N}$, we obtain $X_{k+1}$, $Y_{k+1}$, and $c_{k+1}$ from the updates
    \begin{equation}
        X_{k+1} = \left[
        \begin{smallmatrix}
            \underline{X}_{k} \\
            x_{k}^{\rm{T}}
        \end{smallmatrix}\right],
        \hspace{1mm} Y_{k+1} =  \left[
        \begin{smallmatrix}
            \underline{Y}_{k} \\
            y_{k}^{\rm{T}}
        \end{smallmatrix}\right],
       \hspace{1mm} c_{k+1} =  \left[
        \begin{smallmatrix}
            \big(a_{k}\big)_{[\lnot l_{k}]}\\
            1
        \end{smallmatrix}\right],
        \label{eq:update_c2}
    \end{equation}
    where
    \begin{equation}
        \underline{X}_{k} \triangleq \big(X_{k}\big)_{[\lnot l_{k}]}  \hspace{2mm}, \hspace{2mm}   \underline{Y}_{k} \triangleq \big(Y_{k}\big)_{[\lnot l_{k}]},
        \label{eq:dataset_small}
    \end{equation}   
    \begin{equation}
        (a_{k})_{[i]}\triangleq
        \begin{cases}
            (c_{k})_{[i]},  & i\neq j_{k},\\
            0,              & i=j_{k},
        \end{cases}
        \label{eq:update_c1}
    \end{equation}
    \begin{equation}
        j_{k} \triangleq \argmin_{j \in \{i \hspace{1mm}|\hspace{1mm} (c_{k})_{[i]}=0\}} \big|\big(\Omega(X_{k})^{-1}Q(x_{k},X_{k})\big)_{[j]}\big|,
        \label{eq:smallest_weight}
    \end{equation}
    \begin{equation}
        l_{k} \triangleq \argmax_{l \in \{i\hspace{1mm}|\hspace{1mm} (c_{k})_{[i]}=1\}\cup\{j_{k}\}} \big(P(X_{k})1_{p}\big)_{[l]}.
        \label{eq:largest_correlation}
    \end{equation}

    For each $k$, we note that $(c_{k})_{[i]}=0$ implies that the $i$th row of $X_{k}$ and $Y_{k}$ is data aimed at $(i)$ improving the local model, and $(c_{k})_{[i]}=1$ implies that the $i$th row of $X_{k}$ and $Y_{k}$ is data aimed at $(ii)$ improving the nonlocal model.

    It follows from \cref{eq:dataset_small,eq:update_c2,eq:update_c1} that at each update, the newest data $(x_{k},y_{k})$ is add to improve the local model and the $l_{k}$ row of $X_{k}$ and $Y_{k}$ is removed from the data. To understand how $l_{k}$ is selected, note that it follows from \eqref{eq:GP_stacked_mean} that $\Omega(X_{k})^{-1}Q(x_{k},X_{k})\in \mathbb{R}^{p}$ weights each column of $Y$ in the estimate of $\mu$. Thus, it follows from \cref{eq:smallest_weight} that $j_{k}\in\mathcal{G}_{k}$ is the index out of the $p_{\rm{l}}$ data aimed at $(i)$ improving the local model that has the smallest weight (i.e., least impact) on the estimate $\mu(x,X,Y)$ at the current state $x_{k}$. Hence, data with index $j_{k}$ is removed from $p_{\rm{l}}$ data to make room for new data $(x_{k},y_{k})$. Next, data with index $j_{k}$ is group with the $p_{\rm{g}}$ data aimed at $(ii)$ improving the global model, and we pick one data (i.e., index $l_{k}$) to remove. Note that $l_{k}$ is selected using \eqref{eq:largest_correlation}, and it follows from \eqref{eq:GP_P_def} that $(P(X_{k})1_{p})_{[l]}$ is a measure of how correlated the $l$th data is to the rest of the data set. Hence, $l_{k}$ is select to remove the data that is most correlated.
    
   Finally, we let $\mu_{k}:\mathbb{R}^{n} \rightarrow \mathbb{R}^{n} $, $\sigma_{k}:\mathbb{R}^{n} \rightarrow \mathbb{R}$ and $B_{k} \in \mathbb{R}^{n}$ be defined by
    \begin{equation}
        \mu_{k}(x) \triangleq \mu(x,X_{k},Y_{k}), 
        \label{eq:mean_funct}
    \end{equation}
    \begin{equation}
        \sigma_{k}(x) \triangleq \sigma(x,X_{k}), \qquad  B_{k} \triangleq B(X_{k},Y_{k}).
        \label{eq:upper_funct}
    \end{equation}
        
    \subsection{Recursive Update Equations}  \label{subsec:recursive_alg}
    This section presents recursive update equations for $\mu_{k}$, $\sigma_{k}$ and $B_{k}$ so that these need not be computed directly from \cref{eq:mean_funct,eq:upper_funct} at each update.
    Let $\Sigma_{0}=\Omega(X_{0})^{-1}$, $\vartheta_{0} = \Omega(X_{0})^{-1}Y_{0}$, and $\varsigma_{0}=P(X_{0})1_{p}$. Then, $\Sigma_{k+1} \in \mathbb{R}^{p\times p}$, $\vartheta_{k+1} \in \mathbb{R}^{p\times n}$ and $\varsigma_{k+1} \in \mathbb{R}^{p}$  are computed recursively as
    \begin{gather}
        \Sigma_{k+1} =
        \begin{bmatrix} \underline{\Sigma}_{k}+\frac{\zeta_{k}\zeta_{k}^{\rm{T}}}{\tau_{k}} & -\frac{\zeta_{k}}{\tau_{k}} \\
            -\frac{\zeta_{k}^{\rm{T}}}{\tau_{k}}  &  \frac{1}{\tau_{k}}
        \end{bmatrix}, \label{eq:omega_large}\\
        \vartheta_{k+1} =
        \begin{bmatrix}
            \underline{\vartheta}_{k}+\frac{\zeta_{k}\underline{\mu}_{k}^{\rm{T}}}{\tau_{k}}-\frac{\zeta_{k}y_{k}^{\rm{T}}}{\tau_{k}} \\
            -\frac{\underline{\mu}_{k}^{\rm{T}}}{\tau_{k}}+ \frac{y_{k}^{\rm{T}}}{\tau_{k}}
        \end{bmatrix},
        \label{eq:recursive_vartheta}\\
        \varsigma_{k+1} = \left [
        \begin{smallmatrix}
            \big(\varsigma_{k}-Q\big(\big(X_{k}\big)_{[l_{k}]}^{\rm{T}},X_{k}\big)\big)_{[\lnot l_{k}]}+Q(x_{k},\underline{X}_{k}) \\
            Q(x_{k},X_{k})^{\rm{T}} 1_{p}
        \end{smallmatrix}, \right ]
        \label{eq:extended_varsigma}
        \end{gather}
        where
    \begin{gather}
        \underline{\Sigma}_{k} \triangleq \big(\Sigma_{k}\big)_{[\lnot l_{k},\lnot l_{k}]}-\frac{\big(\Sigma_{k}\big)_{[\lnot l_{k},l_{k}]}\big(\Sigma_{k}\big)_{[\lnot l_{k},l_{k}]}^{\rm{T}}}{\big(\Sigma_{k}\big)_{[l_{k},l_{k}]}},
        \label{eq:omega_small}\\
        \tau_{k}\triangleq q(x_{k},x_{k})-Q(x_{k},\underline{X}_{k})^{\rm{T}}\zeta_{k}+\rho^2,
        \label{eq:tau}\\
        \zeta_{k} \triangleq \underline{\Sigma}_{k}Q(x_{k},\underline{X}_{k}),
        \label{eq:zeta}\\
        \underline{\vartheta}_{k} \triangleq \big(\vartheta_{k}\big)_{[\lnot l_{k}]}-\frac{1}{{\big(\Sigma_{k}\big)_{[l_{k},l_{k}]}}}
        \big(\Sigma_{k}\big)_{[\lnot l_{k},l_{k}]}
        \big(\vartheta_{k}\big)_{[l_{k}]},
        \label{eq:new_vartheta}\\
        \underline{\mu}_{k} \triangleq \underline{\vartheta}_{k}^{\rm{T}}Q(x_{k},\underline{X}_{k}).
        \label{eq:mean_intermediate}
    \end{gather}

    The next result shows that $\Sigma_{k}$, $\vartheta_{k}$, and $\varsigma_{k}$ are recursive computations of $\Omega(X_{k})^{-1}$, $\Omega(X_{k})^{-1}Y_{k}$, and $P(X_{k})1_{p}$.
    The proof is omitted due to space.

    \begin{proposition}
        \rm{For all $k \in \mathbb{N}$, $\Sigma_{k}=\Omega(X_{k})^{-1}$, $\vartheta_{k}=\Omega(X_{k})^{-1}Y_{k}$, and $\varsigma_{k}=P(X_{k})1_{p}$. 
        }
        \label{prop:prop2}
    \end{proposition}

    Proposition \ref{prop:prop2} shows that $\Omega(X_{k})^{-1}$ can be computed recursively using \cref{eq:omega_large,eq:omega_small,eq:tau,eq:zeta}, which does not require matrix inversion. Thus, using Proposition \ref{prop:prop2} and \cref{eq:mean_funct,eq:upper_funct}, it follows that $\mu_{k}$, $\sigma_{k}$, $B_{k}$ can be efficiently computed as
    \begin{equation}
        \mu_{k}(x)=\vartheta_{k}^{\rm{T}}Q(x,X_{k}),
        \label{eq:mean_recursive}
    \end{equation}
    \begin{equation}
        \sigma_{k}(x) = \sqrt{q(x,x)-Q(x,X_{k})^{\rm{T}}\Sigma_{k}Q(x,X_{k})},
        \label{eq:std_recursive}
    \end{equation}
    \begin{equation}
        B_{k} \triangleq \left [
        \begin{smallmatrix}
            \sqrt{b^{2}-(\vartheta_{k}^{\rm{T}}Y_{k})_{[1,1]}+p}\\
            \vdots \\
            \sqrt{b^{2}-(\vartheta_{k}^{\rm{T}}Y_{k})_{[n,n]}+p}
        \end{smallmatrix} \right ],
        \label{eq:B_recursive}
    \end{equation}
    where $X_{k}$, $Y_{k}$, $c_{k}$ are updated using \cref{eq:update_c2,eq:dataset_small,eq:update_c1} and in place of \cref{eq:smallest_weight,eq:largest_correlation}, we use
    \begin{equation}
        j_{k} = \argmin_{j \in \{i \hspace{1mm}|\hspace{1mm} (c_{k})_{[i]}=0\}} \Big|\Big(\Sigma_{k}Q(x_{k},X_{k})\Big)_{[j]}\Big|,
        \label{eq:smallest_weight_recursive}
    \end{equation}
    \begin{equation}
        l_{k} = \argmax_{l \in \{i\hspace{1mm}|\hspace{1mm} (c_{k})_{[i]}=1\}\cup\{j_{k}\}} \big(\varsigma_{k}\big)_{[l]}.
        \label{eq:largest_correlation_recursive}
    \end{equation}

We note from \cref{eq:omega_large,eq:recursive_vartheta,eq:extended_varsigma,eq:omega_small,eq:tau,eq:zeta,eq:new_vartheta,eq:mean_intermediate,eq:mean_recursive,eq:std_recursive,eq:B_recursive,eq:smallest_weight_recursive,eq:largest_correlation_recursive} that this recursive algorithm has time complexity $\mathcal{O}(p^{2})$.
In the next section, \cref{eq:mean_recursive,eq:std_recursive,eq:B_recursive} are incorporated into the safety filter to ensure safety guarantees.

    \section{Safe and Optimal Control}

Let $\xi: [0, \infty) \rightarrow [0,1]$ be a nondecreasing and continuously differentiable function such that for all $t \in (-\infty,0]$,  $\xi(t)=0$ and for all $t \in [1,\infty)$, $\xi(t)=1$. 
The following example provides one possible choice for $\xi$.

\begin{example}\label{ex:xi}\rm
Let $\displaystyle \eta \geq 1$, and consider
		\begin{equation}
			\xi(t) \triangleq
			\begin{cases} 
				0, & t \in (-\infty,0),\\
				\displaystyle \eta t-\frac{\sin{2\pi\eta t}}{2\pi}, & t \in[0,\frac{1}{\eta}],\\
				1, & t \in (\frac{1}{\eta},\infty).  
			\end{cases}
			\label{eq:xit}
		\end{equation}
		\label{ex:Nt}
\end{example}


Define $\mathcal{T}_{k} \triangleq [kT_{\rm{s}}, (k+1)T_{\rm{s}})$ and let $\varphi_{k}: \mathbb{R}^{n} \rightarrow [0,\infty)$ be defined by $\varphi_{k}(x)\triangleq B_{k}\sigma_{k}(x)$.
For all $k \in \mathbb{N}$ and all $t \in \mathcal{T}_{k}$, let $\mu:\mathcal{T}_{k} \times \mathbb{R}^{n}\rightarrow \mathbb{R}^{n}$ be given by
	\begin{equation}
		\mu(t,x) \triangleq
		\xi\bigg( \frac{t-t_{k}}{T_{\rm{s}}}\bigg) \mu_{k}(x) + \bigg[1-\xi\bigg( \frac{t-t_{k}}{T_{\rm{s}}}\bigg)\bigg] \mu_{k-1}(x),
		\label{eq:mutx}
	\end{equation}
 and let $\varphi: \mathcal{T}_{k} \times \mathbb{R}^{n} \rightarrow [0,\infty)$ be given by
	\begin{equation}
		\varphi (t,x) \triangleq
		\xi\bigg( \frac{t-t_{k}}{T_{\rm{s}}}\bigg)\varphi_{k}(x)+\bigg[1-\xi\bigg( \frac{t-t_{k}}{T_{\rm{s}}}\bigg)\bigg]\varphi_{k-1}(x),
		\label{eq:ctx}
	\end{equation}
where $\varphi_{-1}(x) \triangleq \varphi_{0}(x)$ and $\mu_{-1}(x) \triangleq \mu_{0}(x)$.
Note that $\varphi$ and $\mu$ are continuously differentiable functions constructed from the sequences $\varphi_{k}(x)$ and $\mu_{k}(x)$.

        Let $\alpha:\mathbb{R} \rightarrow \mathbb{R}$ be locally Lipschitz and nondecreasing such that $\alpha(0)=0$, and consider the state-constraint function $\psi:\mathbb{R}^{n} \times \mathbb{R}^{n} \times \mathbb{R}^{n} \times \mathbb{R}^{m} \times \mathbb{R} \rightarrow \mathbb{R}$ be defined by
        \begin{align}
            \psi(x,&\hat{\mu},\hat{\varphi},\hat{u},\hat{\delta})\triangleq L_{f}\psi_{d-1}(x)+L_{g}\psi_{d-1}(x)\hat{u}+\hat{\delta}\psi_{d-1}(x) \nn \\
            & +\frac{\partial \psi_{d-1}(x)}{\partial x}\hat{\mu}- \bigg|\frac{\partial \psi_{d-1}(x)}{\partial x}\bigg|\hat{\varphi}+\alpha(\psi_{d-1}(x)),
            \label{eq:CBF_constraint}
        \end{align}
        where $\hat{u}$ is the control variable and $\hat{\delta}$ is a slack variable.

        Define
        \begin{equation}
            \psi_{*}(x,\hat{u},\hat{\delta}) \triangleq \psi(x,w(x),0,\hat{u},\hat{\delta}),
        \end{equation}
        and note that $\psi_{*}(x(t),\hat{u},\hat{\delta}) \geq 0$ is a CBF-based state constraint for \eqref{eq:dyn} that guarantees for all $t\geq 0$, $x(t)\in \bar C$. However, $\psi_{*}$ depends on the unknown function $w$. The next result shows that if for all $i\in \{1,2,...,n\}$, $(\hat{\varphi})_{[i]} \geq |(w-\hat{\mu})_{[i]}|$, then $\psi$ is a lower bound for $\psi_{*}$. Since Proposition \ref{prop:deterministic_error_bound} implies that for all $i\in\{1,2,...,n\}$, $(\varphi(t,x))_{[i]} \geq |(w(x(t))-\mu(t,x))_{[i]}|$, then $\mu$ and $\varphi$ can be used in \eqref{eq:CBF_constraint} to obtain a constraint sufficient for $\psi_{*}(x,\hat{u},\hat{\delta})\geq0$. The proof is similar to that \cite[Proposition 4]{gutierrez2024adaptive} and is omitted for space.
                
        \begin{proposition} \label{prop:constraint_bound}
            \rm{Let $x \in \mathbb{R}^{n}, \hat{u} \in \mathbb{R}^{m}$ and $\hat{\delta} \in \mathbb{R}$. Then, the following hold:}
            \begin{enumerate}
                \item Let $\hat{\mu}\in \mathbb{R}^{n}$ and for all $i \in \{1,2,...,n\}$, let $(\hat{\varphi})_{[i]} \geq \Big|(\hat{\mu}-w(x))_{[i]}\Big|$. Then, $\psi(x,\hat{\mu},\hat{\varphi},\hat{u},\hat{\delta}) \leq \psi_{*}(x,\hat{u},\hat{\delta})$. \label{prop:prop3:a}
                \item For all $t \geq 0$, $\psi(x,\mu(t,x),\varphi(t,x),\hat{u},\hat{\delta}) \leq \psi_{*}(x,\hat{u},\hat{\delta})$. \label{prop:prop3:b}
            \end{enumerate}
        \end{proposition}

        Next, let $\beta>0$, and consider the cost function $\bar{J}:\mathbb{R}^{n} \times \mathbb{R}^{n} \times \mathbb{R}^{m} \times \mathbb{R}  \rightarrow \mathbb{R}$ given by
        \begin{equation}
            \bar{J}(x,\mu,\hat{u},\hat{\delta})\triangleq J(x,\mu,\hat{u})+\frac{\beta}{2}\hat{\delta}^{2},
            \label{eq:sec_cost_func}
        \end{equation}
        which is equal to \eqref{eq:main_cost_func} plus a term that weights the slack variable $\hat{\delta}$. For each $t \geq 0$, the objective is to synthesize $(\hat{u},\hat{\delta})$ that minimizes $\bar{J}(x(t),\mu,\hat{u},\hat{\delta})$ subject to the CBF state constraint $\psi(x(t),\mu,\varphi,\hat{u},\hat{\delta}) \geq 0$, where $\mu$ and $\varphi$ are given by \eqref{eq:GP_stacked_mean} and \eqref{eq:GP_stacked_std}, respectively.

        For all $(x,\mu,\varphi)\in \mathbb{R}^{n} \times \mathbb{R}^{n} \times \mathbb{R}^{n}$, the minimizer of $\bar{J}(x,\mu,\hat{u},\hat{\delta})$ subject to $\psi(x,\mu,\varphi,\hat{u},\hat{\delta})$ can be obtained from the first-order necessary conditions for optimality. The first-order necessary conditions yield the control $u_{*}:\mathbb{R}^{n} \times \mathbb{R}^{n} \times \mathbb{R}^{n}  \rightarrow \mathbb{R}^{m}$ defined by
        \begin{equation}
            u_{*}(x,\mu,\varphi) \triangleq u_{\rm{d}}(x,\mu)+\lambda_{*}(x,\mu,\varphi)H(x,\mu)^{-1}L_{g}\psi_{d-1}(x)^{\rm{T}},
            \label{eq:optim_u}
        \end{equation}
        and the slack variable $\delta_{*}: \mathbb{R}^{n} \times \mathbb{R}^{n} \times \mathbb{R}^{n} \rightarrow \mathbb{R}$ given by
        \begin{equation}
            \delta_{*}(x,\mu,\varphi) \triangleq \beta^{-1}\psi_{d-1}(x)\lambda_{*}(x,\mu,\varphi),
            \label{eq:optim_slack}
        \end{equation}
        where $\lambda_{*}: \mathbb{R}^{n} \times \mathbb{R}^{n} \times \mathbb{R}^{n} \rightarrow \mathbb{R}$ is defined by
       \begin{equation}
		\lambda_{*}(x,\mu,\varphi) \triangleq
		\begin{cases} 
			\displaystyle -\frac{\omega(x,\mu,\varphi)}{\varepsilon(x,\mu)}, & \omega(x,\mu,\varphi)<0,\\
			0, & \omega(x,\mu,\varphi) \geq 0,
		\end{cases}
		\label{eq:optim_lambda}
	\end{equation}
	and $\varepsilon: \mathbb{R}^{n} \times \mathbb{R}^{n} \rightarrow \mathbb{R}$ and $\omega: \mathbb{R}^{n} \times \mathbb{R}^{n} \times \mathbb{R}^{n} \rightarrow     \mathbb{R}$ are given by
        \begin{equation}
            \begin{aligned}
                \varepsilon(x,\mu) &\triangleq L_{g}\psi_{d-1}(x)H(x,\mu)^{-1}L_{g}\psi_{d-1}(x)^{\rmT}\\
                & \qquad +\beta^{-1} \psi_{d-1}(x)^{2},
            \end{aligned}
        \label{eq:optim_r}
        \end{equation}
        \begin{equation}
    	\omega(x,\mu,\varphi) \triangleq \psi(x,\mu,\varphi,u_{\rm{d}}(x,\mu),0).
    	\label{eq:optim_omega}
        \end{equation}

        The next result shows that $(u_{*}(x,\mu,\varphi),\delta_{*}(x,\mu,\varphi))$ is the unique global minimizer of $\bar{J}(x,\mu,\hat{u},\hat{\delta})$ subject to $\psi(x,\mu,\varphi,\hat{u},\hat{\delta})$. The proof is similar to that of \cite[Theorem 1]{safari2024time} and is omitted for space. 

    \begin{theorem}\label{th:cf_optimality}\rm 
        Assume Assumption~\ref{assum:assump2} is satisfied. 
        Let $(x,\mu,\varphi) \in \mathbb{R}^{n}\times \mathbb{R}^{n}\times \mathbb{R}^{n}$, and let $(\hat u, \hat \delta) \in \mathbb{R}^{m}\times \mathbb{R}$ be such that $(\hat u, \hat \delta) \neq (u_{*}(x,\mu,\varphi),\delta_{*}(x,\mu,\varphi)$ and $\psi(x,\mu,\varphi,\hat{u},\hat{\delta}) \geq 0$. 
        Then, $\bar{J}(x,\mu,\hat{u},\hat{\delta})>\bar{J}(x,\mu,u_{*}(x,\mu,\varphi),\delta_{*}(x,\mu,\varphi)).$
    \end{theorem}
        	
        The following theorem is the main result on satisfaction of the state constraint despite the model uncertainty. 
        This result follows from standard CBF analysis techniques (e.g., \cite{ames2016control}) and is omitted for space.

        \begin{theorem}\label{th:main}\rm 
        Consider \eqref{eq:dyn}, where Assumptions~\ref{assum:input}, ~\ref{assum:assump2} and ~\ref{assum:RKHS_norm} are satisfied. 
        Let $u=u_{*}$, where $u_{*}$ is given by \cref{eq:optim_u,eq:optim_slack,eq:optim_lambda,eq:optim_r,eq:optim_omega}, where $\mu$ and $\varphi$ are given by \cref{eq:omega_large,eq:recursive_vartheta,eq:extended_varsigma,eq:omega_small,eq:tau,eq:zeta,eq:new_vartheta,eq:mean_intermediate,eq:mean_recursive,eq:std_recursive,eq:B_recursive,eq:smallest_weight_recursive,eq:largest_correlation_recursive,eq:mutx,eq:ctx}.
        Assume that $h_{0}^\prime$ is locally Lipschitz. 
        Then, for all $x_0 \in \bar C$, the following statements hold: 
        \begin{enumerate}
            \item There exists a maximum value $t_{\rm m} (x_0 ) \in (0 ,\infty]$ such that \eqref{eq:dyn} with $u = u_*$ has a unique solution on $[0, t_\rm (x_0))$.

            \item For all $t \in [0, t_\rm (x_0))$, $x(t) \in \bar C$.
        
            \item Assume the maximum interval of existence and uniqueness is $t_\rm (x_0) =\infty$. 
            Then, for all $t \ge 0$, $x(t) \in \bar C$.
        
        \end{enumerate}
        \end{theorem}
        
        \Cref{th:cf_optimality,th:main} demonstrate that the control \cref{eq:optim_u,eq:optim_slack,eq:optim_lambda,eq:optim_r,eq:optim_omega} satisfies the state constraint that for all $t \ge 0$, $x(t) \in \SC_0$, and yields a control that is optimal with respect to a constraint optimization that aims to obtain a control as close as possible to $u_\rmd$ subject to the state constraint. 

\section{Inverted Pendulum}

    Consider the pendulum modeled by \eqref{eq:dyn}, where
    \begin{equation}
    f(x) =
    \left[
    \begin{array}{c}
        \dot{\gamma} \\
        f_{2}(x)
    \end{array}
    \right],
    \hspace{1mm}
    g(x)=
    \left[
    \begin{array}{c}
        0\\
        \frac{1}{mL^{2}} 
    \end{array}
    \right],
    \hspace{1mm}
    w(x)=
    \left[
    \begin{array}{c}
        0 \\
        w_{2}(x)
    \end{array}
    \right],
    \nonumber
    \end{equation}
    where $
             w_{2}(x) \triangleq  \frac{1}{mL^{2}}(-k_{1} \gamma -k_{2}\gamma^{3} -k_{3}\tanh{\frac{\dot{\gamma}}{\epsilon_{1}}} -k_{4}\dot{\gamma} -k_{5}\dot{\gamma}^{2}\tanh{\frac{\dot{\gamma}}{\epsilon_{2}}} )$, 
        $f_{2}(x) \triangleq \frac{a_\rmg}{L}\sin{\gamma}$, 
         $x = \left[ \gamma \quad \dot{\gamma} \right]^\rmT$,
    $\gamma$ is the position, $\dot{\gamma}$ is the velocity, $a_\rmg$ is gravity, $\epsilon_{1}=\epsilon_{2} =2$, $m=0.5$ kg and $L=0.15$ m, $k_{1}=0.5$ N.m/rad, $k_{2}=0.35$ $\rm{N.m/rad^{3}}$, $k_{3}=0.15$ N.m, $k_{4}=0.5$ N.m.s, $k_{5}=0.25$ $\rm{N.m.s^{2}/rad^{2}}$, and $x(0)=[0.1745 \hspace{2mm} 0]^{\rm{T}}$. 
    The first two terms of $w_{2}$ represent linear and cubic restitution forces, while the last three terms represent coulomb friction, viscous friction and drag force.

    We implement \cref{eq:omega_large,eq:recursive_vartheta,eq:extended_varsigma,eq:omega_small,eq:tau,eq:zeta,eq:new_vartheta,eq:mean_intermediate,eq:mean_recursive,eq:std_recursive,eq:B_recursive,eq:smallest_weight_recursive,eq:largest_correlation_recursive,eq:mutx,eq:ctx}, where $\rho=1$, $(\mu_{0}(x(0)))_{[2]}=0$, $b=100$,  $p=100$, $p_{\rm{l}}=50$, $T_{\rm{s}}=1$ ms, $\eta=10$, the kernel function is $ q(x_{i},x_{j})=100\exp\big(-0.5||x_{i}-x_{j}||_{2}^{2}\big)$,
    and $X_{0}$ and $Y_{0}$ have all their rows filled with $x_{0}$ and $y_{0}$, respectively. Finally,  $(\varphi_{0}(x(0)))_{[2]}=\sqrt{k(x(0),x(0))}\sqrt{b^2+p}$.
    
    Let $\displaystyle \gamma_{\rm{d}}(t)\triangleq-0.99(\pi/4)\cos{0.5t}$ be the desired angular position. Define the error $e \triangleq \gamma-\gamma_{\rm{d}}$ and
    \begin{equation}
        u_{\rm{d}0}(\hat{\mu}) \triangleq mL^{2}\bigg[-f_{2}-(\hat{\mu})_{[2]}+\ddot{\gamma}_{\rm{d}}-K_{1}e-K_{2}\dot{e}\bigg],
        \nonumber
    \end{equation}
    where $K_{1}=25$ and $K_{2}=50$.

    The desired control is defined by
    \begin{equation}
        u_{\rm{d}}(\hat{\mu}) \triangleq 
        \begin{cases} 
           \displaystyle  u_{\rm{d0}}(\hat{\mu})     ,                                                     & F_{\rm{max}} > |u_{\rm{d0}}(\hat{\mu})|, \\
           \displaystyle \frac{F_{\rm{max}}}{|u_{\rm{d0}}(\hat{\mu})|}u_{\rm{d0}}(\hat{\mu})      ,        & F_{\rm{max}} \leq |u_{\rm{d0}}(\hat{\mu})|,
       \end{cases}         
        \nn
    \end{equation}
    where $F_{\rm{max}}=0.35$ N.
    If $u(t)=u_{\rm{d}}(w(x(t)))$, then $\ddot{e}+K_{1}\dot{e}+K_{2}e=0$, which implies that $\lim_{t \rightarrow \infty} e(t)=0$ and $\lim_{t\rightarrow \infty} \dot{e}(t)=0$.
    
    The safe set is given by \eqref{set:safe_set}, where $\psi_{0}(x) = (\frac{\pi}{4})^{2}-\gamma^{2}$.
    We implement the control \cref{eq:optim_u,eq:optim_lambda,eq:optim_slack,eq:optim_omega,eq:optim_r}, where $H=2$, $\beta=200$ and $\alpha_{0}=200$ and $\alpha=20$. The control is updated at 1000 Hz using a zero-order hold.
    
    To examine the impact of the adaptive estimate $\mu$ and the adaptive bound $\varphi$, we consider 3 cases: 
    
    \begin{enumerate}[label=\alph*)]
        \item Adaptive estimate $\mu$ and adaptive bound $\varphi$ are used in CBF constraint and desired control (i.e., $\psi=\psi(x,\mu,\varphi,\hat{u},\hat{\delta})$, and $u_{\rm{d}}=u_{\rm{d}}(\mu)$).
        \item Adaptive estimate $\mu$ and adaptive bound $\varphi$ are used in CBF constraint (i.e., $\psi=\psi(x,\mu,\varphi,\hat{u},\hat{\delta})$), but the desired control uses initial estimate (i.e., $u_{\rm{d}}=u_{\rm{d}}((\mu_{0}(x(0)))_{[2]})$).
        \item Initial estimate $(\mu_{0}(x(0)))_{[2]}$ and initial bound $(\varphi_{0}(x(0)))_{[2]}$ are used in CBF constraint (i.e., $\psi=\psi(x,(\mu_{0}(x(0)))_{[2]},(\varphi_{0}(x(0)))_{[2]},\hat{u},\hat{\delta})$), but the desired control uses the adaptive estimate (i.e., $u_{\rm{d}}=u_{\rm{d}}(\mu)$).
    \end{enumerate}
    \vspace{-0mm}
    Proposition \ref{prop:constraint_bound} implies that all 3 cases satisfy the state constraint. However, Cases 2 and 3 use the initial estimates, which can be conservative and may lead to worse performance.
    \vspace{-5mm}
   \begin{figure}[ht!]
   \begin{subfigure}[t]{0.37\linewidth} 
       \includegraphics[trim={0cm 0cm 0cm 0cm},clip, width=1.02\linewidth]{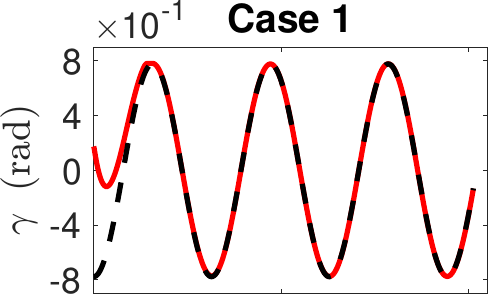}
   \end{subfigure}
    \hspace{-0.15cm}
   \begin{subfigure}[t]{0.295\linewidth} 
       \includegraphics[trim={0cm 0cm 0cm 0cm},clip, width=1.05\linewidth]{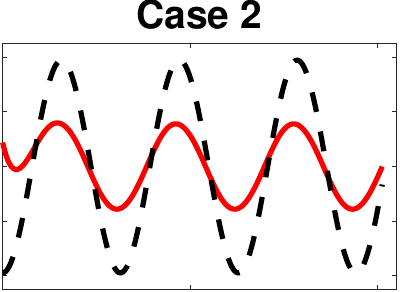}
   \end{subfigure}
   \hspace{-0.075Cm}
   \begin{subfigure}[t]{0.295\linewidth} 
       \includegraphics[trim={0cm 0cm 0cm 0cm},clip, width=1.06\linewidth]{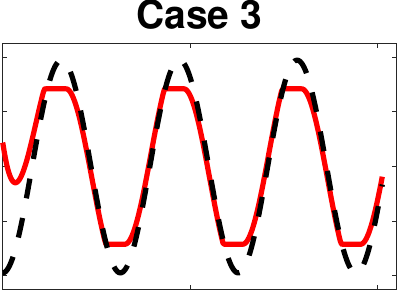}
    \end{subfigure}
   \\
   \begin{subfigure}[t]{0.37\linewidth} 
       \includegraphics[trim={0cm 0cm 0cm -0.00cm},clip, width=1.02\linewidth]{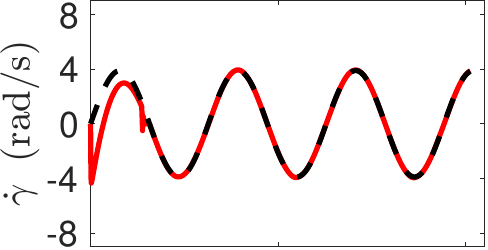}
   \end{subfigure}
   \hspace{-0.15cm}
   \begin{subfigure}[t]{0.295\linewidth} 
       \includegraphics[trim={0cm 0cm 0cm 0cm},clip, width=1.05\linewidth]{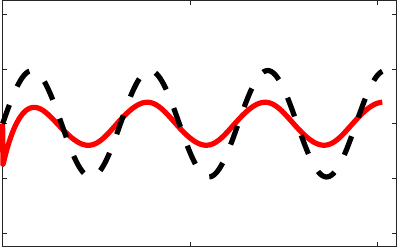}
   \end{subfigure}
   \hspace{-0.075cm}
   \begin{subfigure}[t]{0.295\linewidth} 
       \includegraphics[trim={0cm 0cm 0cm 0cm},clip, width=1.06\linewidth]{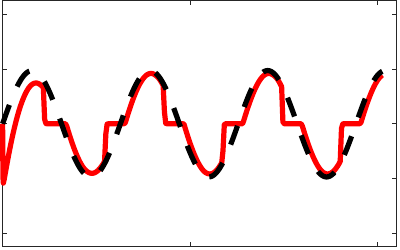}
   \end{subfigure}
    \\   
   \begin{subfigure}[t]{0.37\linewidth} 
       \includegraphics[trim={0cm 0cm 0cm -0.00cm},clip, width=1.02\linewidth]{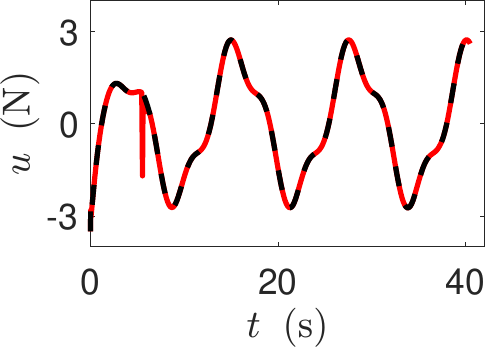}
   \end{subfigure}
   \hspace{-0.18cm}
   \begin{subfigure}[t]{0.295\linewidth} 
       \includegraphics[trim={0cm 0cm 0cm 0cm},clip, width=1.06\linewidth]{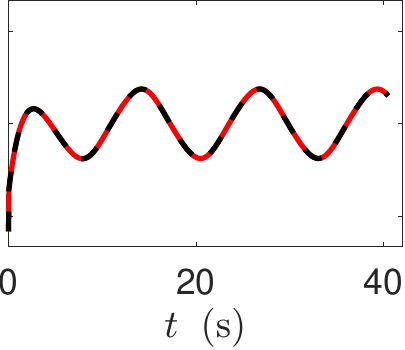}
   \end{subfigure}
   \hspace{-0.075cm}
   \begin{subfigure}[t]{0.295\linewidth} 
       \includegraphics[trim={0cm 0cm 0cm 0cm},clip, width=1.06\linewidth]{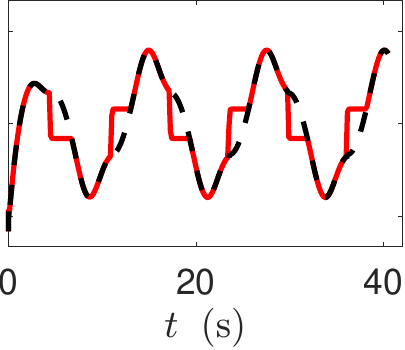}
   \end{subfigure}
   \caption{$\gamma$, $\dot{\gamma}$ and $u$ for Cases 1, 2 and 3, Note that $\gamma_{\rm{d}}$, $\dot{\gamma}_{\rm{d}}$ and $u_{\rm{d}}$ are shown with dashed lines.}
   \label{fig:pend:states}
   \end{figure}

   \vspace{-7mm}
   \begin{figure}[ht!]
   \begin{subfigure}[t]{0.37\linewidth} 
       \includegraphics[trim={0cm 0cm 0cm 0cm},clip, width=1.05\linewidth]{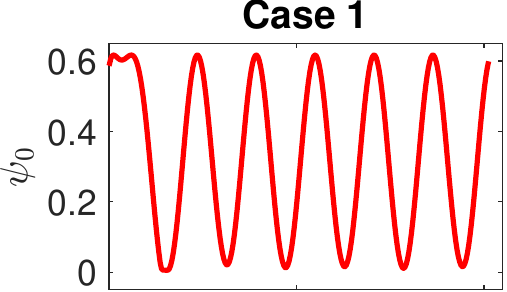}
   \end{subfigure}
    \hspace{-0.05cm}
   \begin{subfigure}[t]{0.295\linewidth} 
       \includegraphics[trim={0cm 0cm 0cm 0cm},clip, width=1.05\linewidth]{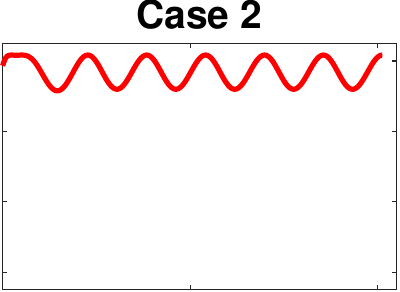}
   \end{subfigure}
   \hspace{-0.05Cm}
   \begin{subfigure}[t]{0.295\linewidth} 
       \includegraphics[trim={0cm 0cm 0cm 0cm},clip, width=1.05\linewidth]{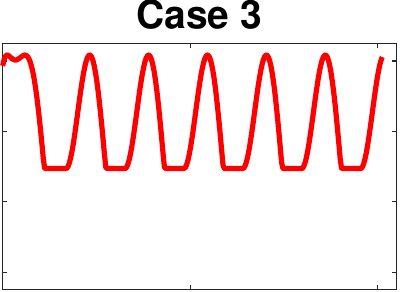}
    \end{subfigure}
   \\
   \begin{subfigure}[t]{0.37\linewidth} 
       \includegraphics[trim={0cm 0cm 0cm -0.00cm},clip, width=1.05\linewidth]{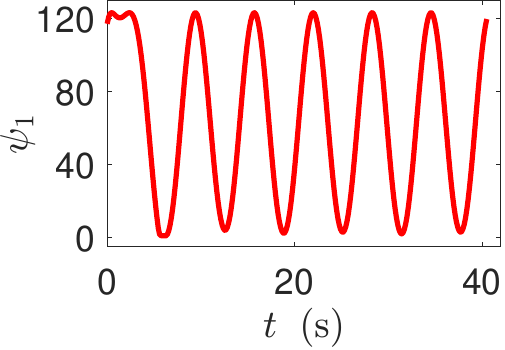}
   \end{subfigure}
   \hspace{-0.1cm}
   \begin{subfigure}[t]{0.295\linewidth} 
       \includegraphics[trim={0cm 0cm 0cm 0cm},clip, width=1.055\linewidth]{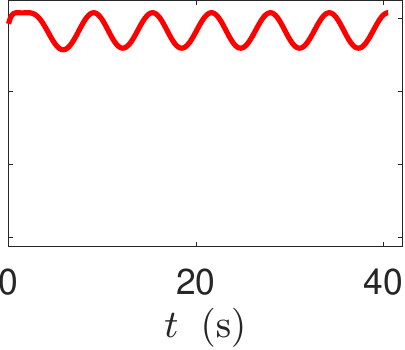}
   \end{subfigure}
   \hspace{-0.075cm}
   \begin{subfigure}[t]{0.295\linewidth} 
       \includegraphics[trim={0cm 0cm 0cm 0cm},clip, width=1.05\linewidth]{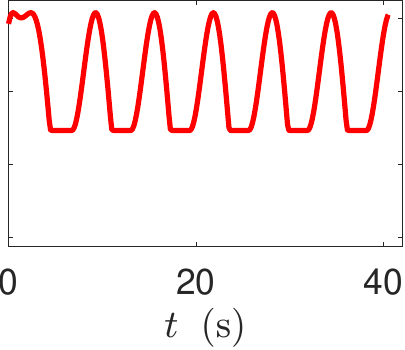}
   \end{subfigure}
   \caption{$\psi_{0}$ and $\psi_{1}$ for Cases 1, 2 and 3.}
   \label{fig:pend:barriers}
   \end{figure}

\begin{figure}[ht!]
    \begin{subfigure}[t]{0.37\linewidth} 
        \includegraphics[trim={0cm 0cm 0cm -0.00cm},clip, width=1.0\linewidth]{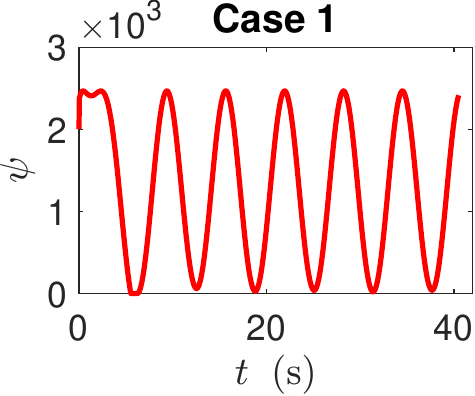}
    \end{subfigure}
    \hspace{-0.25cm}
    \begin{subfigure}[t]{0.295\linewidth} 
        \includegraphics[trim={0cm 0cm 0cm 0cm},clip, width=1.0675\linewidth]{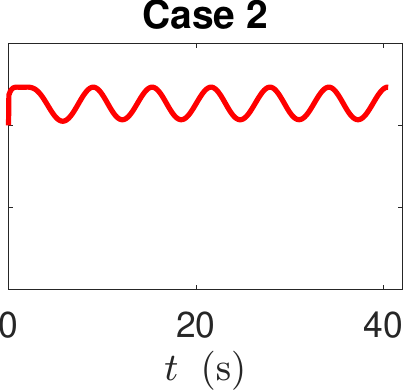}
    \end{subfigure}
    \hspace{-0.05cm}
    \begin{subfigure}[t]{0.295\linewidth} 
        \includegraphics[trim={0cm 0cm 0cm 0cm},clip, width=1.0675\linewidth]{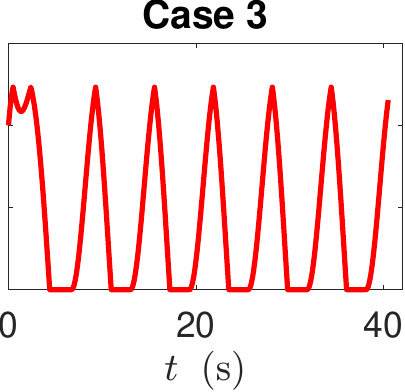}
    \end{subfigure}
    \caption{$\psi$ for Cases 1, 2 and 3.}
    \label{fig:pend:const}
   \end{figure}

\begin{figure}[ht!]
   \begin{subfigure}[t]{0.37\linewidth} 
       \includegraphics[trim={0cm 0cm 0cm 0cm},clip, width=1.02\linewidth]{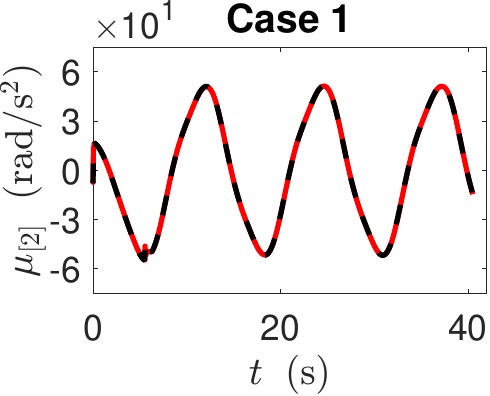}
   \end{subfigure}
    \hspace{-0.15cm}
   \begin{subfigure}[t]{0.295\linewidth} 
       \includegraphics[trim={0cm 0cm 0cm 0cm},clip, width=1.06\linewidth]{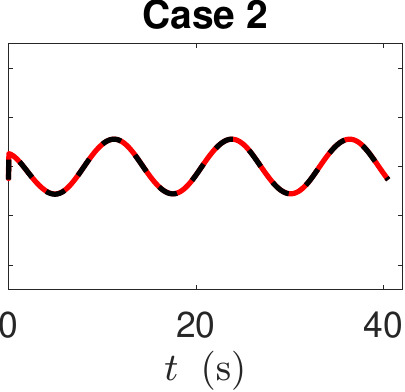}
   \end{subfigure}
   \hspace{-0.05Cm}
   \begin{subfigure}[t]{0.295\linewidth} 
       \includegraphics[trim={0cm 0cm 0cm 0cm},clip, width=1.0675\linewidth]{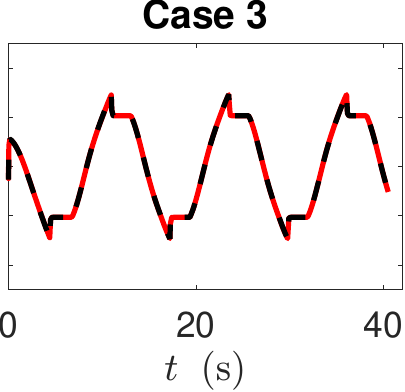}
    \end{subfigure}
   \caption{$\mu_{[2]}$ for Cases 1, 2 and 3. Note that $w_{[2]}$ is shown using dashed line.}
   \label{fig:pend:est}
   \end{figure}

\begin{figure}[ht!]
   \begin{subfigure}[t]{0.37\linewidth} 
       \includegraphics[trim={0cm 0cm 0cm 0cm},clip, width=1.05\linewidth]{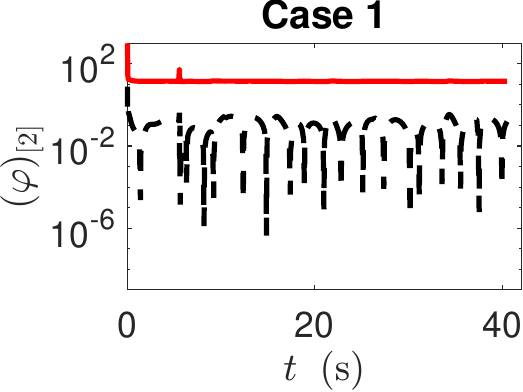}
   \end{subfigure}
    \hspace{0.04cm}
   \begin{subfigure}[t]{0.295\linewidth} 
       \includegraphics[trim={0cm 0cm 0cm 0cm},clip, width=1.03\linewidth]{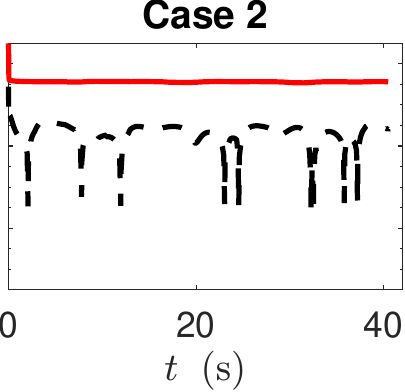}
   \end{subfigure}
   \hspace{-0.05cm}
   \begin{subfigure}[t]{0.295\linewidth} 
       \includegraphics[trim={0cm 0cm 0cm 0cm},clip, width=1.03\linewidth]{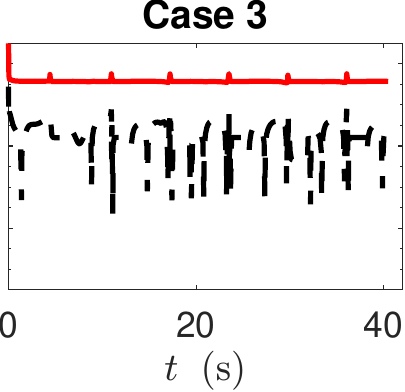}
    \end{subfigure}
   \caption{$(\varphi)_{[2]}$ for Cases 1, 2 and 3. Note that $|(\mu-w)_{[2]}|$ is shown with dashed lines.}
   \label{fig:pend:est_error}
   \end{figure}

    \vspace{-3mm}
   Figure \ref{fig:pend:states} shows $\gamma$, $\dot{\gamma}$ and $u$, where $x(0)=[0.1745 \hspace{2mm} 0]^{\rm{T}}$. In Case 1 it is shown that $\displaystyle \lim _{t \rightarrow \infty} \gamma(t) =\gamma_{\rm{d}}$ and $\displaystyle \lim _{t \rightarrow \infty} \dot{\gamma}(t)=\dot{\gamma}_{\rm{d}}$. In Case 2 it is shown that $\gamma$ and $\dot{\gamma}$ are not following $\gamma_{\rm{d}}$ and $\dot{\gamma}_{\rm{d}}$, respectively. In Case 3 it is shown that $\gamma$ and $\dot{\gamma}$ follow $\gamma_{\rm{d}}$ and $\dot{\gamma}_{\rm{d}}$, except for the region near $\gamma=\pm \frac{\pi}{4}$ rad.
    Figures \ref{fig:pend:barriers} and \ref{fig:pend:const} show that for all $t \geq 0$, $\psi_{0}$, $\psi_{1}$ and $\psi$ are nonnegative for Cases 1, 2 and 3. Furthermore, in Case 1 the safety constraint is not activated in steady-state. In Case 2, the safety constraint is positive for all $t \geq 0$. In Case 3, the safety constraint is periodically activated because of its conservativeness. This explain why in Case 3 $\gamma$ and $\dot{\gamma}$ does not follow $\gamma_{\rm{d}}$ and $\dot{\gamma}_{\rm{d}}$ near to $\gamma=\pm \frac{\pi}{4}$ rad.
   Figure \ref{fig:pend:est} shows that $\mu_{[2]}$ follows $w_{[2]}$ for all three Cases. Figure \ref{fig:pend:est_error} shows that for all $t \geq 0$, the estimation error $|(\mu-w)_{[2]}|$ is bounded by $(\varphi)_{[2]}$.
   
    \vspace{-2mm}
   
    \vspace{-7mm}

\vspace{-7mm}

  \vspace{-2mm} 
\section{Nonholonomic Ground Robot}
    Consider the nonholonomic differential drive mobile robot modeled by \eqref{eq:dyn}, where
    \begin{equation}
        f(x)=
        \left[\hspace{-1mm}
        \begin{array}{c}
            v\cos{\gamma}-l_{\rm{d}}\omega\sin{\gamma} \\
            v\sin{\gamma}+l_{\rm{d}}\omega\cos{\gamma} \\
            \omega \\
            0 \\
            0
        \end{array}
        \hspace{-1mm}\right],  \hspace{1mm}
        g(x)=
        \left[\hspace{-1mm}
        \begin{array}{cc}
            0 & 0 \\
            0 & 0 \\
            0 & 0\\
            g_{1}(x) & g_{2}(x)
        \end{array}
        \hspace{-1mm}\right],
        \nonumber
    \end{equation} 
    \begin{equation}
    w(x)=
    \left[\hspace{-1mm}
    \begin{array}{c}
        0 \\
        0 \\
        0 \\
        w_{4}(x) \\
        w_{5}(x)
    \end{array}
    \hspace{-1mm}\right], \hspace{0mm}
    x = \left[
        \begin{array}{c}
            q_{\rm{x}}\\
            q_{\rm{y}}\\
            \gamma\\
             v \\
             \omega
        \end{array}
        \right], \hspace{0mm}
        u = \left[\hspace{-1mm}
        \begin{array}{c}
            u_{\rm{r}}\\
            u_{\rm{l}}
        \end{array}
        \hspace{-1mm}\right],
    \nonumber
    \end{equation}
    where 
    \begin{equation}
        \begin{aligned}
             w_{4}(x) \hspace{-1mm} \triangleq &  \displaystyle -2\Big(\frac{k_{\rm{b}}k_{\rm{m}}}{mrR_{\rm{a}}}+\frac{\epsilon}{mr}\Big) \Big(v+a_{1}v^{2}\tanh(2.5v)\Big)  \\
             & \qquad -\kappa g \sin(\gamma),\\
        \end{aligned}
    \nonumber
    \end{equation}

    \begin{equation}
    \begin{aligned}
        w_{5}(x) \hspace{-1mm} \triangleq &  -\Big(\frac{k_{\rm{b}}k_{\rm{m}}l^2}{Ir^2R_{\rm{a}}}+\frac{l\epsilon}{Ir^2}\Big)\Big(\omega+a_{2}\omega^{2}\tanh(2.5\omega)\Big),\\
    \end{aligned}
    \nonumber
    \end{equation}
    
    \begin{equation}
        g_{1}(x)\triangleq
        \left[\hspace{-2mm}
        \begin{array}{cc}
            \displaystyle \frac{k_{\rm{m}}}{mrR_{\rm{a}}} & \displaystyle \frac{k_{\rm{m}}l}{IrR_{\rm{a}}}
        \end{array}
        \hspace{-2mm}\right]^{\rm{T}}\hspace{-2mm}, \hspace{2mm}
        g_{2}(x)\triangleq
        \left[\hspace{-2mm}
        \begin{array}{cc}
            \displaystyle \frac{k_{\rm{m}}}{mrR_{\rm{a}}} & \displaystyle -\frac{k_{\rm{m}}l}{IrR_{\rm{a}}}
        \end{array}
        \hspace{-2mm}\right]^{\rm{T}},
        \nonumber
    \end{equation} 
     and $[q_{\rm{x}} \hspace{2mm} q_{\rm{y}}]^{\rm{T}}$ denote the position of the tip of the robot in an orthogonal coordinate frame, $\gamma$ is the direction of the velocity vector, $v$ and $\omega$ are the velocity and angular velocity, $u_{\rm{r}}$ and $u_{\rm{l}}$ are the voltage of each motor, $k_{\rm{m}}=0.1$ N-m/Amp is the torque constant, $r=0.1$ m is the wheel radius, $l=0.5$ m is the distance between wheels, $l_{\rm{d}}=0.25$ m is the distance from the center of mass to the tip of the vehicle, $R_{\rm{a}}=0.27$ ohms is the armature resistance, $m=10$ kg and $I=0.83$ kg-$\rm{m^{2}}$ are the vehicle mass and inertia, $a_{1}=1$ s/m, $a_{2}=1$ s and $g$ is the gravity. Moreover, $k_{\rm{b}} = 0.0487$ V/(rad/sec) is the back-EMF constants of the motors, $\epsilon = 0.025$ N-m-s is a friction coefficient, and $\kappa = 0.5$ corresponds to an angle of inclination of the ground of $30 ^{\circ}$. This dynamic is based on that in  \cite{anvari2013non}. However, unlike their model, we incorporate a drag damping term in $\dot{v}$ and $\dot{\omega}$, along with a gravitational term in $\dot{v}$ that accounts for ground inclination.

    We implement \cref{eq:omega_large,eq:recursive_vartheta,eq:extended_varsigma,eq:omega_small,eq:tau,eq:zeta,eq:new_vartheta,eq:mean_intermediate,eq:mean_recursive,eq:std_recursive,eq:B_recursive,eq:smallest_weight_recursive,eq:largest_correlation_recursive,eq:mutx,eq:ctx}, where $\rho=0.5$, $(\mu_{0}(x(0)))_{[4]}=0$, $(\mu_{0}(x(0)))_{[5]}=0$, $b= 100$, $p=100$, $p_{\rm{l}}=50$, $T_{\rm{s}}=1$ ms, $\eta=10$, the kernel function is $ q(x_{i},x_{j})=100\exp \big(-0.1\big|\big|x_{i}-x_{j}\big|\big|_{2}^{2}\big)$,
     and $X_{0}$ and $Y_{0}$ have all their rows filled with $x_{0}$ and $y_{0}$, respectively. Finally, $(\varphi_{0}(x(0)))_{[4]}=\sqrt{k(x(0),x(0))}\sqrt{b^2+p}$ and $(\varphi_{0}(x(0)))_{[5]}=\sqrt{k(x(0),x(0))}\sqrt{b^2+p}$.
    
    Similar to  \cite{rabiee2025soft}, we define
    \begin{equation}
        a_{\rm{d}}\triangleq -\big(\mu_{1}+\mu_{2}\big)v-\big(1+\mu_{1}\mu_{2}\big)e_{1}+\frac{\mu_{1}^{2}}{l_{\rm{d}}}e_{2}^{2}, \hspace{5mm} \omega_{\rm{d}}\triangleq -\frac{\mu_{1}}{l_{\rm{d}}}e_{2},
        \nonumber
    \end{equation}
    where
    \begin{equation}
        e_{1}\triangleq (q_{\rm{x}}-q_{\rm{dx}})\cos{\gamma}+(q_{\rm{y}}-q_{\rm{dy}})\sin{\gamma},
        \nonumber
    \end{equation}
    \begin{equation}
        e_{2}\triangleq -(q_{\rm{x}}-q_{\rm{dx}})\sin{\gamma}+(q_{\rm{y}}-q_{\rm{dy}})\cos{\gamma},
        \nonumber
    \end{equation}
     $\mu_{1}=\mu_{2} =0.25$ and $q_{\rm{d}}=[q_{\rm{dx}} \hspace{2mm} q_{\rm{dy}}]^{\rm{T}}$ is the desired position of the tip of the robot.
    Define $e_{\rm{b}} \triangleq  \omega-\omega_{\rm{d}}$. Also, define
    \begin{equation}
        u_{\rm{d1}}(\hat{\mu})  \triangleq -(\hat{\mu})_{[4]} \hspace{2mm} + \hspace{2mm} a_{\rm{d}} - \hspace{2mm} K_{1}v,
        \nonumber
    \end{equation}
    \begin{equation}
        u_{\rm{d2}}(\hat{\mu})  \triangleq -(\hat{\mu})_{[5]} - \frac{\mu_{1}}{l_{\rm{d}}}\dot{e}_{2}-K_{2}e_{\rm{b}},
        \nonumber
    \end{equation}
    where $K_{1}=2$ and $K_{2}=2$.
    Hence, the desired control is $u_{\rm{d}}(\hat{\mu}) \triangleq [ u_{\rm{dr}}(\hat{\mu}) \hspace{2mm} u_{\rm{dl}}(\hat{\mu})]$, where
    \begin{equation}
        u_{\rm{dr}}(\hat{\mu}) \triangleq
        \displaystyle \frac{mrR_{\rm{a}}}{2k_{\rm{m}}}u_{\rm{d1}}(\hat{\mu})+\frac{IrR_{\rm{a}}}{2k_{\rm{m}}l}u_{\rm{d2}}(\hat{\mu}),
        \nonumber
    \end{equation}
    \begin{equation}
        u_{\rm{dl}}(\hat{\mu}) \triangleq
        \displaystyle \frac{mrR_{\rm{a}}}{2k_{\rm{m}}}u_{\rm{d1}}(\hat{\mu})-\frac{IrR_{\rm{a}}}{2k_{\rm{m}}l}u_{\rm{d2}}(\hat{\mu}),
        \nonumber
    \end{equation}
    If $u(t)=u_{\rm{d}}\Big(w(x(t))\Big)$, then $\dot{v}=-K_{1}v+a_{\rm{d}}$ and $\dot{\omega}-\dot{\omega}_{\rm{d}}=-K_{2}(\omega-\omega_{\rm{d}})$, which implies that $\lim_{t \rightarrow \infty} (\omega(t)-\omega_{\rm{d}}(t)) = 0$.

    For all $j\in\{1,2,3,4\}$, let the obstacles be defined by
    \begin{equation}
        \displaystyle
        \phi_{j,0}(x) \triangleq b_{j}\Big(\big(q_{\rm{x}}-c_{j,1}\big)^{2}+\big(q_{\rm{y}}-c_{j,2}\big)^2-R^2\Big),
        \nonumber
    \end{equation}
    where $R=0.6$ m, $b_{1}=1$, $b_{2}=b_{3}=b_{4}=0.5$, $c_{1,1}=0.35$ m, $c_{1,2}=0.7$ m, $c_{2,1}=2.75$ m, $c_{2,2}=1.75$ m, $c_{3,1}=2.5$ m, $c_{3,2}=-0.25$ m, $c_{4,1}=1$ m, $c_{4,2}=2.2$ m. 
    The wall is modelled by
    \begin{equation}
        \phi_{5,0}(x) \triangleq -\frac{1}{\varrho_{5}}\log\Big(\sum_{i=1}^{4}\exp(-\varrho_{5}(h_{\rm{w}}(x))_{[i]})\Big),
        \nn
    \end{equation}
    where $(h_{\rm{w}}(x))_{[1]} \triangleq q_{\rm{x}}+1$, $(h_{\rm{w}}(x))_{[2]} \triangleq 4-q_{\rm{x}}$, $(h_{\rm{w}}(x))_{[3]} \triangleq q_{\rm{y}}+1$, $(h_{\rm{w}}(x))_{[4]}\triangleq 3-q_{\rm{y}}$ and $\varrho_{5}=20$.   
    The bounds on $v$ and $\omega$ are modelled by
    \begin{equation}
        \phi_{6,0}(x) \triangleq 0.5 \Big(1-v^{2}\Big), \hspace{2mm}  \phi_{7,0}(x) \triangleq 0.5 \Big(1-\omega^{2}\Big).
        \nn
    \end{equation}
    Note that $\phi_{6,0}(x)$ and $\phi_{7,0}(x)$ are relative degree 1 whereas $\phi_{1,0}(x)$, ... , $\phi_{5,0}(x)$ are relative degree 2. Thus, for $j\in \{1,2,...,5\}$, define
    \begin{equation}
	\phi_{j,1}(x) \triangleq L_{f}\phi_{j,0}(x)+\alpha_{j,0} \big( \phi_{j,0}(x) \big),
	\nn
    \end{equation}
    where $\alpha_{j,0}=2$.

    The \textit{soft-min} approach presented in \cite{rabiee2024closed} is implemented to compose the multiple CBFs with different relative degrees. 
    This approach uses the the log-sum-exponential soft minimum $\text{softmin}_{\varrho} : \mathbb{R} \times \cdots \times \mathbb{R} \rightarrow \mathbb{R}$ defined by $\text{softmin}_{\varrho}(z_{1},z_{2},\cdots,z_{q}) \triangleq -\frac{1}{\varrho}\log \sum_{i=1}^{q} e^{-\varrho z_{i}}$, where $\varrho>0$.
   The safe set \eqref{set:safe_set} is modeled by the zero-superlevel set of
    \begin{equation}
        \psi_{0}(x)=\text{softmin}_{\varrho}\Big(\phi_{1,1}(x),\phi_{2,1}(x),...,\phi_{6,0}(x),\phi_{7,0}(x)\Big),
        \nn
    \end{equation}
    where $\varrho=20$ and $d=1$.

    We implement the control \cref{eq:optim_u,eq:optim_lambda,eq:optim_slack,eq:optim_omega,eq:optim_r}, where $H=2I_{2}$, $\beta=2$  and $\alpha=1$. The control is updated at $1000$ Hz using a zero-order hold structure.
    The three Cases examined in the first examples are revisited to analyze the impact of the estimate $\mu$ and adaptive bound $\varphi$ in Case 1, and to examine the conservative safety constraint and the degraded performance in Cases 2 and 3.

\begin{figure}[ht!]
        \centering
        \includegraphics[trim={0.12cm 0cm 2.0cm 0.3cm}, clip, width=0.85\linewidth]{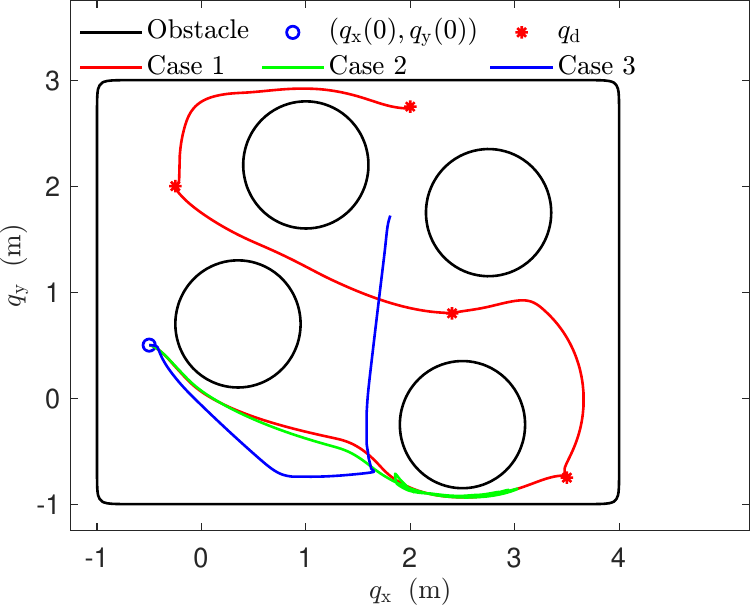}
        \caption{Phase-Portrait of 3 closed-loop trajectories for Cases 1, 2 and 3. The notation $\rm{bd}$ $C_{*}$ denotes the boundary of $C_{*}$.}
        \label{fig:GR:phase_port}
    \end{figure}

\begin{figure}[ht!]
   \begin{subfigure}[t]{0.37\linewidth} 
       \includegraphics[trim={0cm 0cm 0cm 0cm},clip, width=1.02\linewidth]{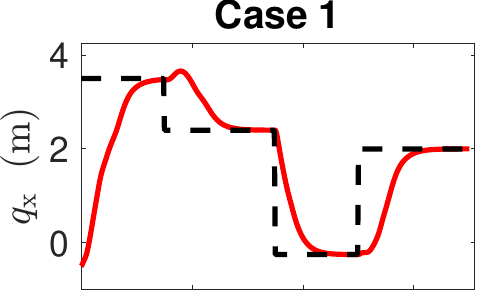}
   \end{subfigure}
    \hspace{-0.15cm}
   \begin{subfigure}[t]{0.295\linewidth} 
       \includegraphics[trim={0cm 0cm 0cm 0cm},clip, width=1.0675\linewidth]{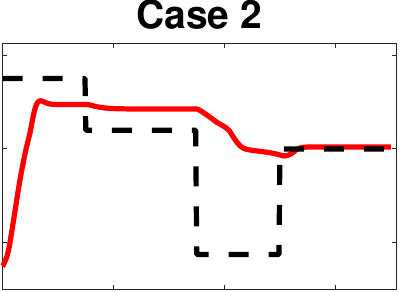}
   \end{subfigure}
   \hspace{-0.05Cm}
   \begin{subfigure}[t]{0.295\linewidth} 
       \includegraphics[trim={0cm 0cm 0cm 0cm},clip, width=1.0675\linewidth]{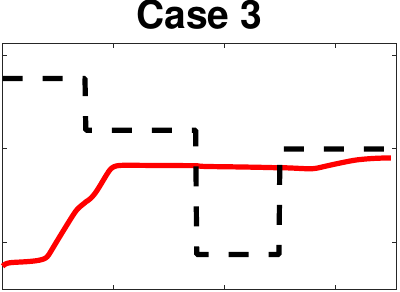}
    \end{subfigure}
   \\
   \begin{subfigure}[t]{0.37\linewidth} 
       \includegraphics[trim={0cm 0cm 0cm -0.00cm},clip, width=1.02\linewidth]{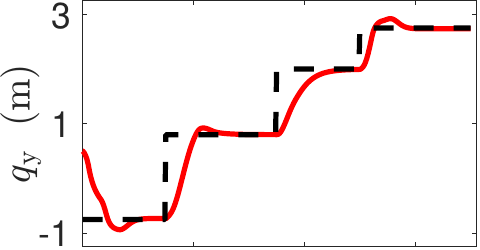}
   \end{subfigure}
   \hspace{-0.15cm}
   \begin{subfigure}[t]{0.295\linewidth} 
       \includegraphics[trim={0cm 0cm 0cm 0cm},clip, width=1.0675\linewidth]{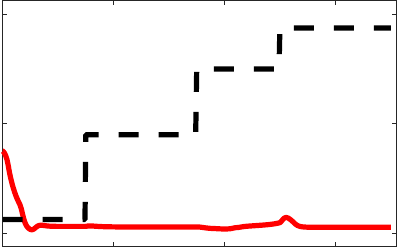}
   \end{subfigure}
   \hspace{-0.05cm}
   \begin{subfigure}[t]{0.295\linewidth} 
       \includegraphics[trim={0cm 0cm 0cm 0cm},clip, width=1.0675\linewidth]{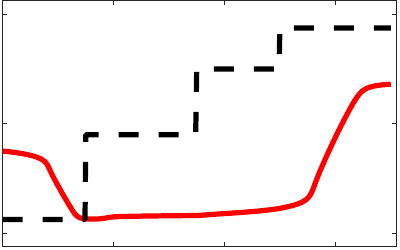}
   \end{subfigure}
    \\   
   \begin{subfigure}[t]{0.37\linewidth} 
       \includegraphics[trim={0cm 0cm 0cm -0.00cm},clip, width=1.02\linewidth]{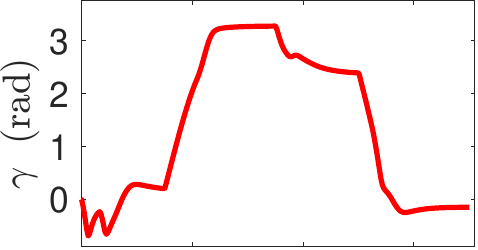}
   \end{subfigure}
   \hspace{-0.15cm}
   \begin{subfigure}[t]{0.295\linewidth} 
       \includegraphics[trim={0cm 0cm 0cm 0cm},clip, width=1.0675\linewidth]{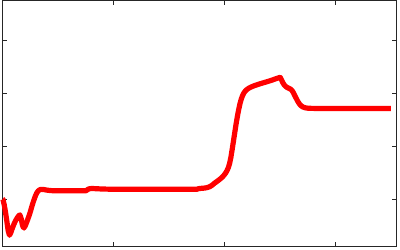}
   \end{subfigure}
   \hspace{-0.05cm}
   \begin{subfigure}[t]{0.295\linewidth} 
       \includegraphics[trim={0cm 0cm 0cm 0cm},clip, width=1.0675\linewidth]{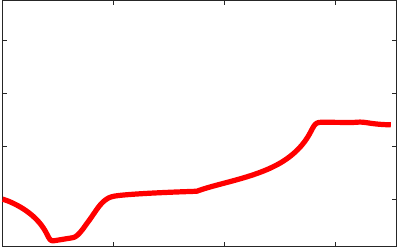}
   \end{subfigure}
  \\
  \begin{subfigure}[t]{0.37\linewidth} 
      \includegraphics[trim={0cm 0cm 0cm -0.0cm},clip, width=1.02\linewidth]{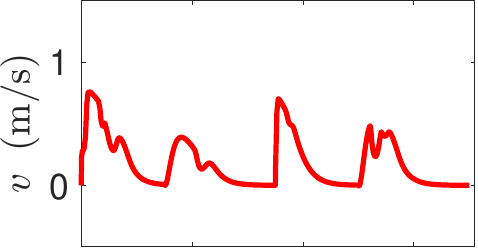}
  \end{subfigure}
   \hspace{-0.15cm}
   \begin{subfigure}[t]{0.295\linewidth} 
       \includegraphics[trim={0cm 0cm 0cm 0cm},clip, width=1.0675\linewidth]{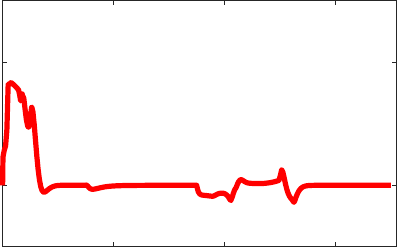}
   \end{subfigure}
   \hspace{-0.05cm}
   \begin{subfigure}[t]{0.295\linewidth} 
       \includegraphics[trim={0cm 0cm 0cm 0cm},clip, width=1.0675\linewidth]{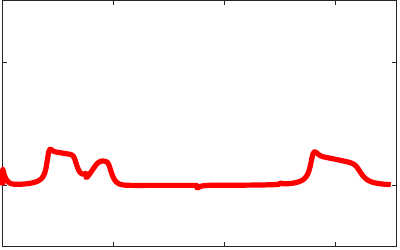}
   \end{subfigure}
    \\
   \begin{subfigure}[t]{0.37\linewidth} 
       \includegraphics[trim={0cm 0cm 0cm -0.0cm},clip, width=1.02\linewidth]{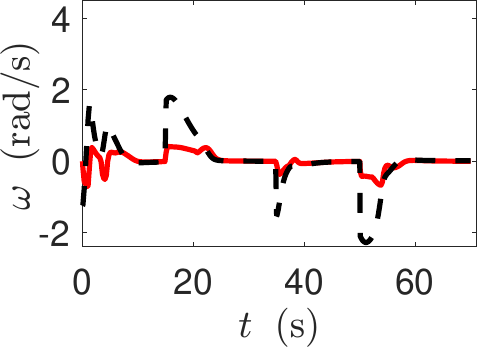}
   \end{subfigure}
   \hspace{-0.18cm}
   \begin{subfigure}[t]{0.295\linewidth} 
       \includegraphics[trim={0cm 0cm 0cm 0cm},clip, width=1.075\linewidth]{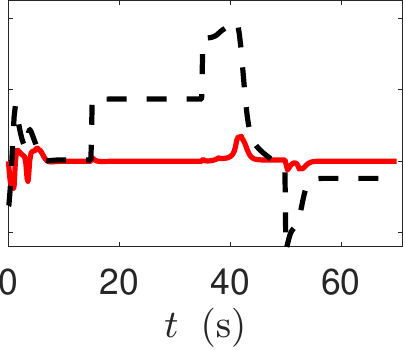}
   \end{subfigure}
   \hspace{-0.05cm}
   \begin{subfigure}[t]{0.295\linewidth} 
       \includegraphics[trim={0cm 0cm 0cm 0cm},clip, width=1.075\linewidth]{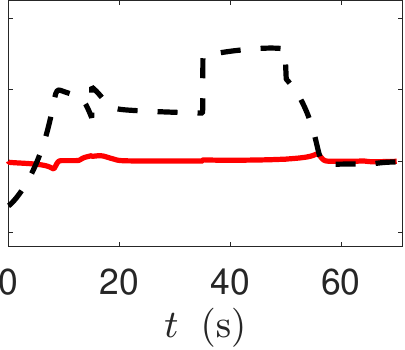}
   \end{subfigure}
   \caption{$q_{\rm{x}}$, $q_{\rm{y}}$, $\gamma$, $v$ and $\omega$ for Cases 1, 2 and 3. Note that $q_{\rm{dx}}$, $q_{\rm{dy}}$ and $\omega_{\rm{d}}$ are shown with dashed lines.}
   \label{fig:GR:states}
   \end{figure}

\begin{figure}[ht!]
   \begin{subfigure}[t]{0.37\linewidth} 
       \includegraphics[trim={0cm 0cm 0cm 0cm},clip, width=1.03\linewidth]{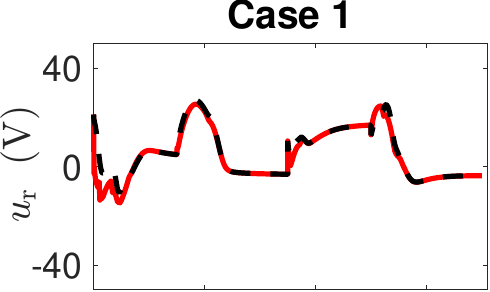}
   \end{subfigure}
   \hspace{-0.13cm}
   \begin{subfigure}[t]{0.295\linewidth} 
       \includegraphics[trim={0cm 0cm 0cm 0cm},clip, width=1.06\linewidth]{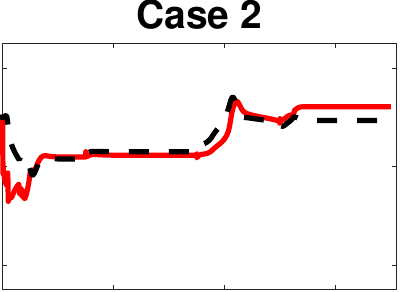}
   \end{subfigure}
   \hspace{-0.075cm}
   \begin{subfigure}[t]{0.295\linewidth} 
       \includegraphics[trim={0cm 0cm 0cm 0cm},clip, width=1.06\linewidth]{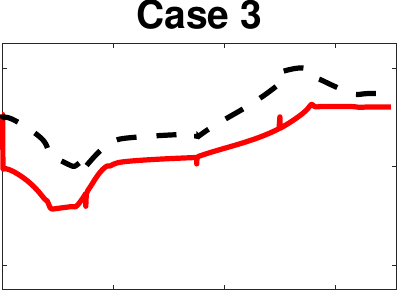}
  \end{subfigure}
\\
   \begin{subfigure}[t]{0.37\linewidth} 
       \includegraphics[trim={0cm 0cm 0cm -0.0cm},clip, width=1.03\linewidth]{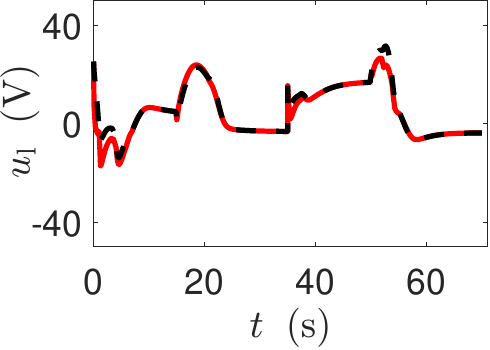}
   \end{subfigure}
   \hspace{-0.16cm}
   \begin{subfigure}[t]{0.295\linewidth} 
       \includegraphics[trim={0cm 0cm 0cm 0cm},clip, width=1.065\linewidth]{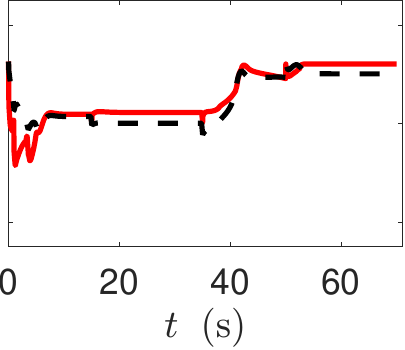}
   \end{subfigure}
   \hspace{-0.075cm}
   \begin{subfigure}[t]{0.295\linewidth} 
       \includegraphics[trim={0cm 0cm 0cm 0cm},clip, width=1.065\linewidth]{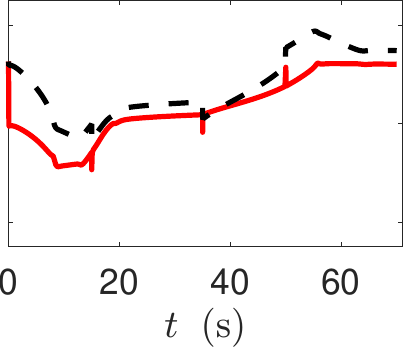}
   \end{subfigure}
\\
   \caption{$u_{\rm{r}}$ and $u_{\rm{l}}$ for Cases 1, 2 and 3. Note that $u_{\rm{dr}}$ and $u_{\rm{dl}}$ are shown with dashed lines.}
   \label{fig:GR:control}
   \end{figure}

\begin{figure}[ht!]
    \begin{subfigure}[t]{0.37\linewidth} 
        \includegraphics[trim={0cm 0cm 0cm 0cm},clip, width=1.02\linewidth]{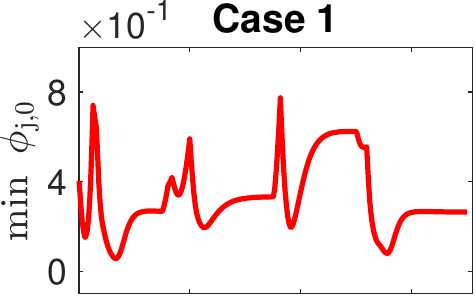}
    \end{subfigure}
    \hspace{-0.15cm}
    \begin{subfigure}[t]{0.295\linewidth} 
        \includegraphics[trim={0cm 0cm 0cm 0.0cm},clip, width=1.075\linewidth]{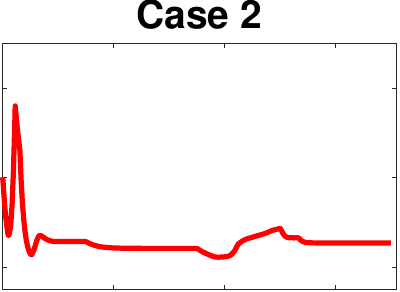}
    \end{subfigure}
    \hspace{-0.025cm}
    \begin{subfigure}[t]{0.295\linewidth} 
        \includegraphics[trim={0cm 0cm 0cm 0.0cm},clip, width=1.07\linewidth]{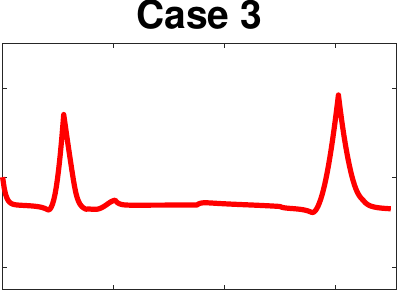}
    \end{subfigure}
\\ 
    \begin{subfigure}[t]{0.37\linewidth} 
        \includegraphics[trim={0cm 0cm 0.cm 0.0cm},clip, width=1.02\linewidth]{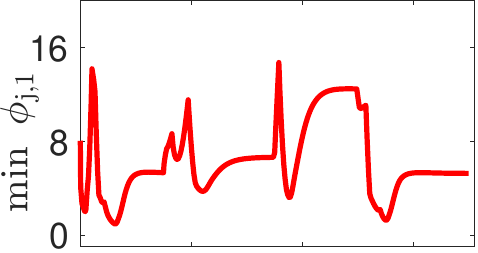}
    \end{subfigure}
    \hspace{-0.15cm}
    \begin{subfigure}[t]{0.295\linewidth} 
        \includegraphics[trim={0.cm 0cm 0cm -0.0cm},clip, width=1.077\linewidth]{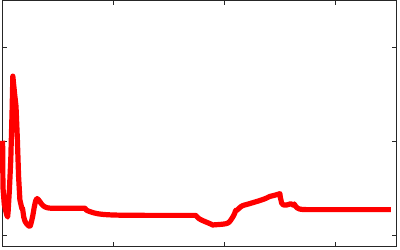}
    \end{subfigure}
    \hspace{-0.025cm}
    \begin{subfigure}[t]{0.295\linewidth} 
        \includegraphics[trim={-0cm 0cm 0cm -0.0cm},clip, width=1.08\linewidth]{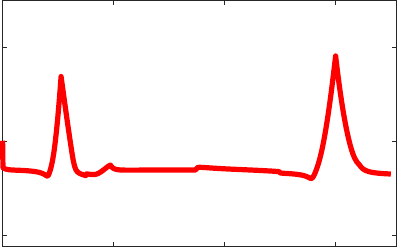}
    \end{subfigure}
\\
    \begin{subfigure}[t]{0.37\linewidth} 
        \includegraphics[trim={0cm 0cm 0.cm 0.0cm},clip, width=1.02\linewidth]{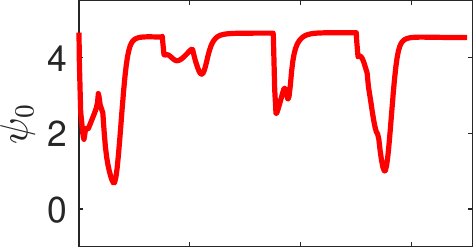}
    \end{subfigure}
    \hspace{-0.15cm}
    \begin{subfigure}[t]{0.295\linewidth} 
        \includegraphics[trim={0.cm 0cm 0cm -0.0cm},clip, width=1.08\linewidth]{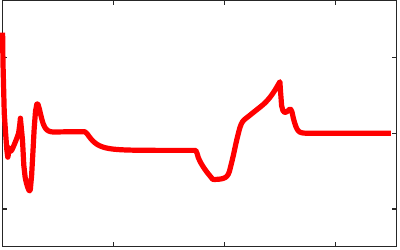}
    \end{subfigure}
    \hspace{-0.025cm}
    \begin{subfigure}[t]{0.295\linewidth} 
        \includegraphics[trim={-0cm 0cm 0cm -0.0cm},clip, width=1.08\linewidth]{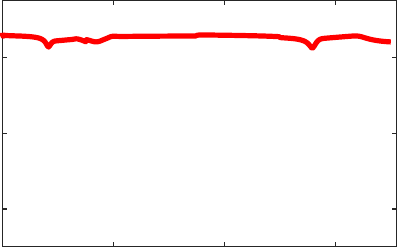}
    \end{subfigure}
\\
    \begin{subfigure}[t]{0.37\linewidth} 
        \includegraphics[trim={0cm 0cm 0.cm 0.0cm},clip, width=1.02\linewidth]{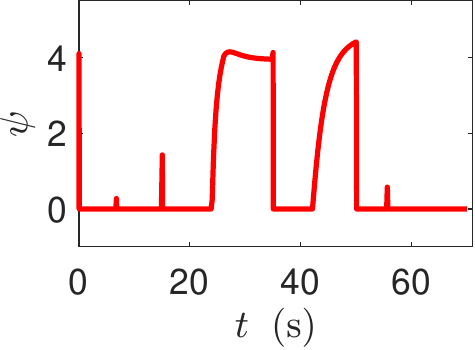}
    \end{subfigure}
    \hspace{-0.19cm}
    \begin{subfigure}[t]{0.295\linewidth} 
        \includegraphics[trim={0.cm 0cm 0cm -0.0cm},clip, width=1.092\linewidth]{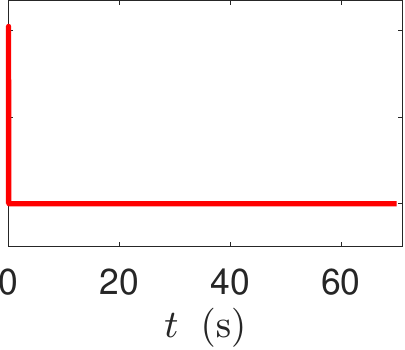}
    \end{subfigure}
    \hspace{-0.025cm}
    \begin{subfigure}[t]{0.295\linewidth} 
        \includegraphics[trim={-0cm 0cm 0cm -0.0cm},clip, width=1.085\linewidth]{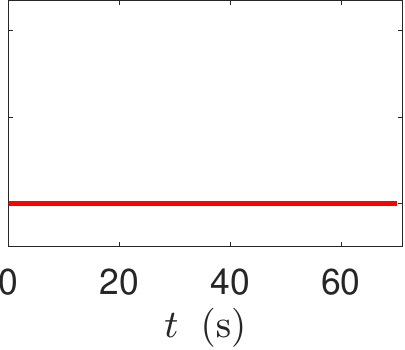}
    \end{subfigure}
    \caption{$\min \phi_{j,0}$, $\min \phi_{j,1}$, $\psi_{0}$ and $\psi$ for Cases 1, 2 and 3.}
    \label{fig:GR:Barrier_Functions}
    \end{figure}

\begin{figure}[ht!]
    \begin{subfigure}[t]{0.37\linewidth} 
        \includegraphics[trim={0cm 0cm 0cm 0cm},clip, width=1.03\linewidth]{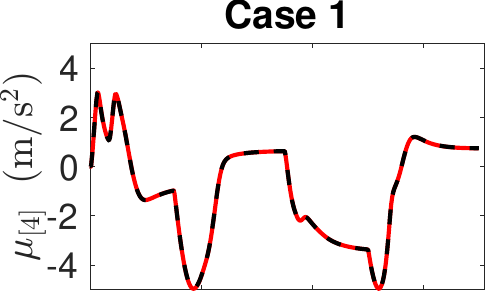}
    \end{subfigure}
    \hspace{-0.12cm}
    \begin{subfigure}[t]{0.29\linewidth} 
        \includegraphics[trim={0cm 0cm 0cm 0.0cm},clip, width=1.08\linewidth]{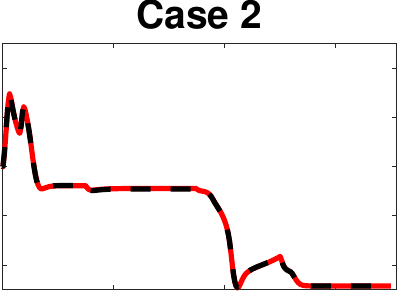}
    \end{subfigure}
    \hspace{-0.01cm}
    \begin{subfigure}[t]{0.29\linewidth} 
        \includegraphics[trim={0cm -0.04cm 0cm 0.0cm},clip, width=1.078 \linewidth]{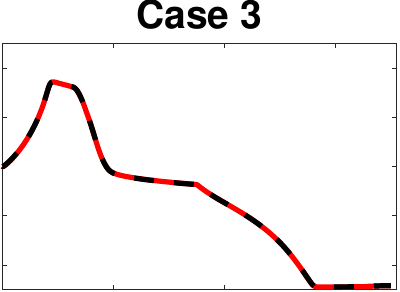}
    \end{subfigure}
\\
    \begin{subfigure}[t]{0.37\linewidth} 
        \includegraphics[trim={0cm 0cm 0.cm 0.0cm},clip, width=1.03\linewidth]{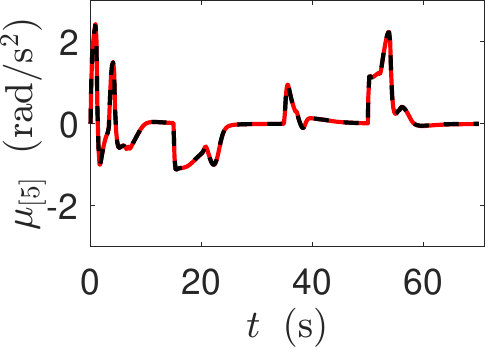}
    \end{subfigure}
    \hspace{-0.16cm}
    \begin{subfigure}[t]{0.29\linewidth} 
        \includegraphics[trim={0.cm 0cm 0cm -0.0cm},clip, width=1.09\linewidth]{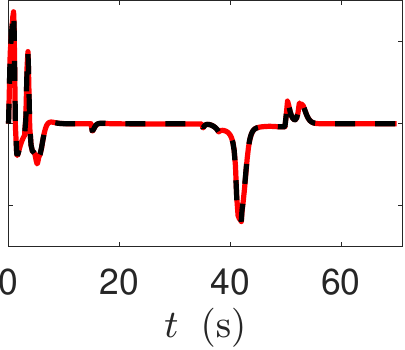}
    \end{subfigure}
    \hspace{-0.01cm}
    \begin{subfigure}[t]{0.29\linewidth} 
        \includegraphics[trim={-0cm 0cm 0cm -0.0cm},clip, width=1.09\linewidth]{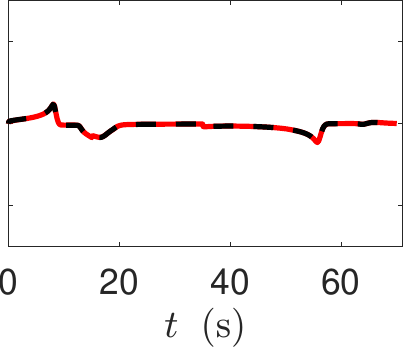}
    \end{subfigure}
        \caption{$\mu_{[4]}$ and $\mu_{[5]}$ for Cases 1, 2 and 3. Note that $w_{[4]}$ and $w_{[5]}$ are shown with dashed lines.}
    \label{fig:GR:est}
    \end{figure}

\begin{figure}[ht!]
   \begin{subfigure}[t]{0.37\linewidth} 
       \includegraphics[trim={0cm 0cm 0cm 0cm},clip, width=1.07\linewidth]{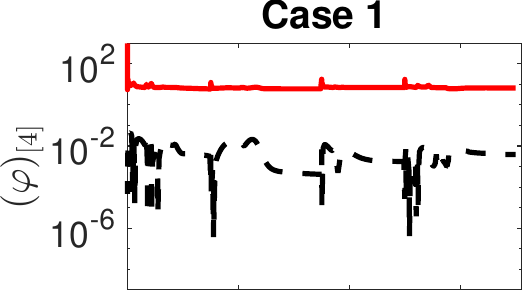}
   \end{subfigure}
   \hspace{0.07cm}
   \begin{subfigure}[t]{0.295\linewidth} 
       \includegraphics[trim={0cm 0cm 0cm 0cm},clip, width=1.03\linewidth]{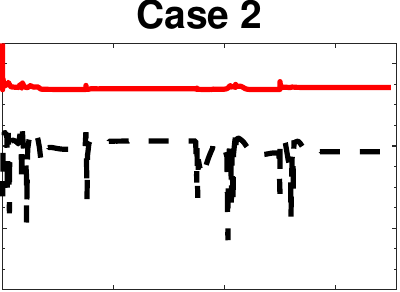}
   \end{subfigure}
   \begin{subfigure}[t]{0.295\linewidth} 
       \includegraphics[trim={0cm 0cm 0cm 0cm},clip, width=1.03\linewidth]{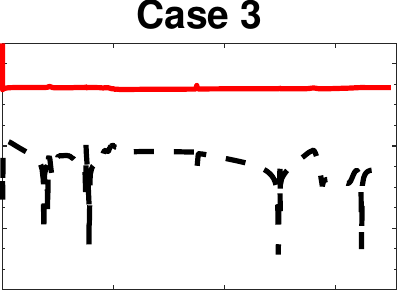}
   \end{subfigure}
    \hspace{-0.12cm}
   \\
   \begin{subfigure}[t]{0.37\linewidth} 
       \includegraphics[trim={0cm 0cm 0cm 0cm},clip, width=1.07\linewidth]{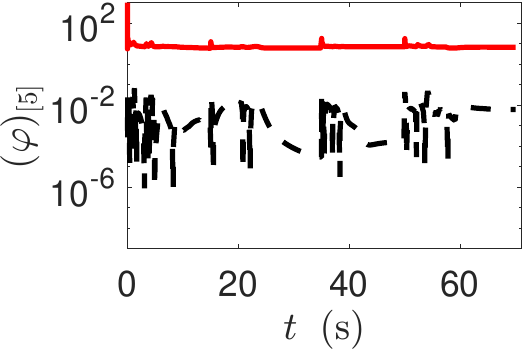}
  \end{subfigure}
   \hspace{0.03cm}
  \begin{subfigure}[t]{0.295\linewidth} 
       \includegraphics[trim={0cm 0cm 0cm 0cm},clip, width=1.035\linewidth]{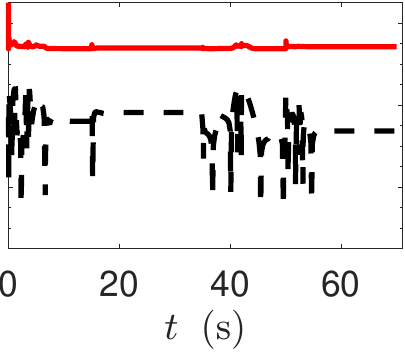}
  \end{subfigure}
  \begin{subfigure}[t]{0.295\linewidth} 
       \includegraphics[trim={0cm 0cm 0cm 0cm},clip, width=1.035\linewidth]{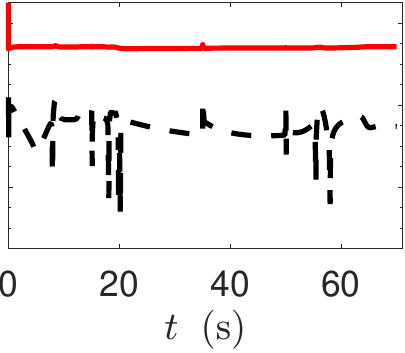}
  \end{subfigure}
    \hspace{-0.1cm}
        \caption{$\big(\varphi\big)_{[4]}$ and $\big(\varphi\big)_{[5]}$ for Cases 1, 2 and 3. Note that $|(\mu-w)_{[4]}|$ and $|(\mu-w)_{[5]}|$ are shown with dashed lines.}
    \label{fig:GR:est_error}
    \end{figure}

    Figure \ref{fig:GR:phase_port} shows the closed-loop trajectories for $x_{0}=[-0.5 \hspace{2mm} 0.5  \hspace{2mm} 0  \hspace{2mm} 0  \hspace{2mm} 0]^{\rm{T}}$, with a sequence of goals shown in figure for each case.
    Figure \ref{fig:GR:states} shows the evolution of the states for the three proposed cases. It is shown that only for Case 1, $q_{\rm{x}}$, $q_{\rm{y}}$ and $\omega$ are driven to $q_{\rm{dx}}$, $q_{\rm{dy}}$ and $\omega_{\rm{d}}$.
   Figure \ref{fig:GR:control} shows the control $u_{\rm{r}}$ and $u_{\rm{l}}$ and the desired control $u_{\rm{dr}}$ and $u_{\rm{dl}}$. Note that in Case 3, $u_{\rm{r}}$ and $u_{\rm{l}}$ does not follow $u_{\rm{dr}}$ and $u_{\rm{dl}}$, respectively.
    Figure \ref{fig:GR:Barrier_Functions} shows that for all three Cases, $\phi_{1,0}$, ... , $\phi_{5,0}$, $\phi_{1,1}$, ... , $\phi_{5,1}$, $\psi_{0}$ and $\psi$ are nonnegative, which implies that all constraints are satisfied. In Case 3, it is shown that the safety filter remains permanently active, which explains the overall performance degradation.
    Figure \ref{fig:GR:est} shows that for all three Cases, $\mu_{[4]}$ and $\mu_{[5]}$ follow $w_{[4]}$ and $w_{[5]}$, respectively.
    Figure \ref{fig:GR:est_error} shows that $\big|\big(\mu-w\big)_{[4]}\big|\leq \big(\varphi\big)_{[4]}$ and $\big|\big(\mu-w\big)_{[5]}\big|\leq \big(\varphi\big)_{[5]}$.

    \bibliographystyle{ieeetr}
    \bibliography{References}

 \end{document}